\documentclass[nonblindrev]{informs3}

\OneAndAHalfSpacedXI 

\usepackage{natbib}
 \bibpunct[, ]{(}{)}{,}{a}{}{,}%
 \def\bibfont{\small}%
 \def\newblock{\ }%
\usepackage{mathtools}
\usepackage{float}
\usepackage{amsmath}
\usepackage{algorithm}
\usepackage{algpseudocode}
\usepackage{graphicx}
\usepackage{array}
\usepackage[table,xcdraw]{xcolor}
\usepackage{tabularx}
\usepackage{booktabs} 
\TheoremsNumberedThrough     
\ECRepeatTheorems

\EquationsNumberedThrough    

\newcommand{\curly}[1]{\left\{#1\right\}}

\newcommand{\pacman}[1]{\(\langle \text{#1} \rangle\)}

\begin{document}


\RUNAUTHOR{Chen and Zaman}

\RUNTITLE{Optimizing Influence Campaigns: Nudging under Bounded Confidence}

\TITLE{Optimizing Influence Campaigns: Nudging Under Bounded Confidence}

\ARTICLEAUTHORS{%
\AUTHOR{Yen-Shao Chen}
\AFF{School of Management, Yale University, New Haven, CT 06511, \EMAIL{yen-shao.chen@yale.edu}} 
\AUTHOR{Tauhid Zaman}
\AFF{School of Management, Yale University, New Haven, CT 06511, \EMAIL{tauhid.zaman@yale.edu}}
} 

\ABSTRACT{
	Influence campaigns in online social networks are often run by organizations, political parties, and nation states to influence large audiences. These campaigns are employed through the use of agents in the network that share persuasive content.  Yet, their impact might be minimal if the audiences remain unswayed, often due to the \textit{bounded confidence} phenomenon, where only a narrow spectrum of viewpoints can influence them.  Here we show  that to persuade under bounded confidence, an agent must \textit{nudge} its targets to gradually shift their opinions.  Using a control theory approach, we show how to construct an agent's nudging policy under the bounded confidence opinion dynamics model and also how to select targets for multiple agents in an influence campaign on a social network.  Simulations on real Twitter networks show that a multi-agent nudging policy can shift the mean opinion, decrease opinion polarization, or even increase it.  We find that our nudging based policies outperform other common techniques that do not consider the bounded confidence effect. Finally, we show how to craft prompts for large language models, such as ChatGPT, to generate text-based content for real nudging policies. This illustrates the practical feasibility of our approach, allowing one to go from mathematical nudging policies to real social media content.
}



\maketitle

%

\section{Introduction}\label{sec:intro}
In the advent of the digital era, social media platforms have risen to become central stages for marketing, promotion, and a broad array of persuasive endeavors. These platforms' extensive reach and the connections they foster provide opportunities for a diverse range of actors including political parties \citep{kinnard_trump_2024, fernandez_feral_2024}, governments \citep{gorska_getting_2022}, and non-governmental organizations \citep{armstrong_digital_2018} to shift public opinion and behavior.  Traditionally, these influence campaigns have utilized messaging which directly advocates for a certain opinion on a given topic.  For instance, in public health, many pro-vaccine campaigns have posted messages highlighting the benefits of being vaccinated \citep{gallagher_rhetorical_2020, hoffman_doctorsspeakup_2021}.  Political campaigns often use ads which directly attack opponents or support a candidate \citep{megerian_harris_nodate}.  Nation states have been suspected to utilize bot accounts on social media to manipulate online discourse and advance an agenda \citep{broniatowski_weaponized_2018, golovchenko_cross-platform_2020}.  However, studies have found that many times this type of direct messaging is less effective \citep{aronson_power_1999, li_self-persuasion_2020}.  There are different explanations for this.  One suggests that people exist in echo chambers where they are only exposed to a single perspective.   This leads them to distrust the information that contradicts their existing beliefs \citep{boutyline_social_2017, nguyen_echo_2020}.  This effect has also been termed \emph{bounded confidence} \citep{hegselmann_opinion_2002}.  Under this model, the range within which individuals are open to modifying their beliefs is limited, suggesting that persuasion outside of this range is likely to be ineffective due to the audience's resistance to changing entrenched views.  An even more extreme behavior is known as the \emph{backfire effect} where messages or content with opposing views actually cause one's initial belief to shift away from the target message \citep{bail2018exposure, pluviano_parents_2019, galasso_positive_2023}.  
These complex models of human behavior necessitate a different approach to persuasion.

Direct persuasion is optimal if people behave according to the classical  DeGroot model \citep{degroot1974reaching}.  In this model, all content exhibits an attractive force, pulling opinions towards their message.  This occurs through the weighted averaging of adjacent opinions in a social network.  However, if people's behavior deviates from the DeGroot model, the direct approach may not have the expected persuasive effect.  This has driven researchers to develop new novel persuasion techniques.  One prominent technique, known as \emph{nudging},  subtly guides people towards making certain decisions by modifying the manner in which information is presented \citep{thaler2021nudge}.  In the field of online misinformation, different types of nudging interventions have been proposed, including accuracy prompts, friction, and social norms \citep{kozyreva_toolbox_2024}.  These nudges provide additional information with the misinformation content aimed at preventing its spread.  

The nudging techniques designed for misinformation can be implemented by a social media platform.  However, if one is conducting an influence campaign then typically the tools available are very different in nature.  One cannot control the manner in which content is presented nor what information is provided with it.  Instead, one can control who is posting the message, either through the recruitment of influential users of the platform or the creation of online agents.  By using messengers that are more aligned with the target audience, the persuasion is done  \emph{in-group}.  It has been found that in-group persuasion based on the race of the messenger is effective at modifying the behavior of users \citep{munger_tweetment_2017}.  This can be viewed as a different type of nudge, where the presentation of information is modified by the choice of messenger.  In a related work, it was found that having the information have some alignment with the target was also effective at modifying behavior \citep{yang2022mitigating}.  This type of nudging where the presented content is close to the target's belief does not require a modification of the messenger or additional information with the content, but a modification of the content itself.  While this is not something a social media platform can do, it is exactly what an agent in an influence campaign can do.  This content based nudging can result in effective persuasion for individuals who display bounded confidence.  In contrast, direct persuasion can result in no opinion change under the bounded confidence model.

Content based nudging is a dynamic persuasion strategy, meaning that the content changes over time.  The content causes the target's opinions to shift, and as they do, the content also shifts to continue pulling the target to the desired opinion.  The nature of the bounded confidence model necessitates this type of dynamic strategy.  This approach is in contrast to direct persuasion where the opinion of the content is fixed.  If the audience did not exhibit bounded confidence and was totally persuadable, then this fixed opinion content strategy would be optimal.  However, under bounded confidence, this strategy will not be effective and a dynamic content strategy is required.


Determining the dynamic content policy typically involves solving for a policy for the content opinion, with the opinion expressed as a numerical value.  However, this is not textual content.  To bridge this gap and actually implement policy, we need to convert a numerical opinion to content on the topic with the given opinion.  Today this is feasible through the use of large language models (LLMs) like ChatGPT \citep{van_dis_chatgpt_2023}. These models allow for the conversion of numerical content policies into tangible, engaging content through the use of clever prompt designs.  By combining analytic content strategies with LLMs, one obtains an end-to-end influence campaign that starts with the social network structure and opinions of the target audience, and ends with targeted content designed to persuade the audience.

\subsection{Our Contributions}
In this work we present a control theory approach to designing influence campaigns in social networks under the bounded confidence model.  We model the opinion dynamics in a social network with a system of differential equations.  We then present algorithms for the two phases of an influence campaign: target selection and content creation.  For target selection, we present a greedy optimization algorithm that allows for the effective identification of targets for multiple agents.  For content creation, we solve for the content's time varying opinion using principles from control theory.  We find that  our resulting policy utilizes content based nudging that shifts the audience opinion gradually over time. We perform simulations on real Twitter network topologies and show that our algorithms are more effective at shifting opinions than standard approaches and scalable to large networks.  With our approach, we are able to shift the mean opinion in a network and also increase or decrease the opinion polarization.  Finally, we present prompt engineering techniques to show how to convert the content opinion policy into social media posts using ChatGPT.

This paper is organized as follows. Section \ref{sec:lit} reviews the relevant literature. We present our social network opinion dynamics model in Section \ref{sec:model}. The optimal control theory formulation for an influence campaign, along with the algorithms we develop to solve for the policy, are presented in Section \ref{sec:control}. Section \ref{sec:results} provides policy intuitions from synthetic networks and demonstrates policy effects on large-scale Twitter networks through simulations. LLM prompt engineering for content creation based on content policies is discussed in Section \ref{sec:llm}. Section \ref{sec:conclusion} concludes the paper.

\section{Literature Review}\label{sec:lit}

Much of the extant literature focuses on modeling the mechanisms of persuasion. One of the earliest models proposed to quantify opinion change is the DeGroot model, which employs linear dynamics to achieve consensus among nodes in a network \citep{degroot1974reaching}. Variations of this model have been developed to account for additional considerations such as the importance of one's prior belief \citep{friedkin_social_1990} and the discrepancy between expressed and private opinions \citep{anderson2019recent}. However, modeling becomes increasingly complex when empirical observations are considered. For instance, opinions can become polarized \citep{reiter-haas_polarization_2023}, persuasion efforts can backfire \citep{nyhan_when_2010,bail2018exposure}, and spillover effects may occur \citep{galasso_positive_2023}, not to mention the influence of echo chambers and filter bubbles \citep{nguyen_echo_2020, mosleh2021shared}. These complexities necessitate nonlinear models to more accurately capture reality. A widely used nonlinear model is the bounded confidence opinion dynamics model \citep{hegselmann_opinion_2002}, which has inspired numerous variants addressing different empirical phenomena \citep{bernardo_bounded_2024}. This model posits that persuasion power is negligible when the opinion gap between the sender and the audience is too large, making it one of the most challenging opinion dynamics models to analyze \citep{hegselmann_opinion_2002}. In this study, we use the bounded confidence model and propose a scalable agent content policy aimed at shifting audience opinions in an influence campaign.

To choose agent content over time, one common approach involves deploying a stubborn agent—a pinning controller or zealot—who maintains a constant opinion to influence and recruit followers \citep{mobilia2007role, masuda_opinion_2015, ghezelbash2019polarization, khalil_zealots_2021, hunter2022optimizing}. Under linear DeGroot dynamics, this results in an opinion consensus at the opinion of the stubborn agent. Yet in reality, we have nonlinear opinion dynamics. Field experiment have also found that changing the agent opinions over time can influence subjects' opinions and mitigate polarization \citep{yang2022mitigating}. Thus, we apply dynamic and moreover, adaptive agent opinions over time. One dynamic agent control policy concentrates persuasion efforts on one individual at a time and persuades that target with utmost strength \citep{hegselmann_optimal_2015}. However, this heuristic, while focusing on immediate impacts, overlooks the longitudinal effects of influence and is not suited for larger networks due to its exhaustive enumeration of potential targets. Additionally, the concept of frequently changing influence targets seems impractical, as social media platforms typically exhibit a consistent follower base. Another dynamic agent control policy applies a similar optimal control approach to ours, yet allows for changing agent targets \citep{kozitsin_optimal_2022}. These methods do not fix agents' targets and are not scalable to large networks.

Choosing a fixed set of agent targets is often achieved through a greedy search algorithm \citep{kempe2003maximizing, hunter2022optimizing}. The greedy algorithm is suitable not only because the influence maximization problem itself is NP-hard \citep{kempe2003maximizing}, but also because under linear dynamics, maximizing the mean is submodular in the number of targets \citep{hunter2022optimizing}. Although we assume nonlinear opinion dynamics, we still apply greedy targeting in this study due to its scalability and effectiveness.


\section{Opinion Dynamics Model}\label{sec:model}
The goal of an influence campaign is to shift opinions in a social network using agents.  In order to determine the agents' actions, we first need a model for the social network, the opinions of users in the network, and how the opinions evolve over time as the users interact with one another. We follow the modeling approach of \citep{chen_shaping_2024}, which we now present in detail.

The social network can be represented as a directed graph $G = (V, E)$, where nodes in the set $V$ correspond to users, and edges in the set $E$ represent the follower relationships between them. This model suits social networks with a follower-following structure, such as Twitter, Instagram, or TikTok. In this graph, a directed edge $(i, j)$ from user $i$ to user $j$ means that user $j$ follows user $i$, indicating that user $j$ will receive content posted by user $i$. The rate at which user $i$ shares content with user $j$ is denoted by $\lambda_{ij}$.


Every user \(i\) possesses a latent opinion \(\theta_i(t)\) that varies over time and is represented by a real number. The opinion embedded in the content shared by a user at any time equals their current latent opinion. While more sophisticated models allow for the content's opinion to equal the latent opinion \(\theta_i(t)\) plus zero-mean noise \citep{hunter2022optimizing}, we will not explore such stochastic extensions here.

The rate of change of the opinion of a user $i$ is given by the sum of contributions from all users that $i$ follows.  Each user $j$ that $i$ follows contributes a term $\lambda_{ji}f(\theta_j-\theta_i)$ to this rate, for some opinion shift function $f$.
We assume the shift function depends only on the difference between the two opinions, which is a standard assumption for most opinion dynamics models \citep{degroot1974reaching, hegselmann_opinion_2002}. The presence of the posting rate indicates that users who post more frequently exert greater influence on the opinions of their followers.  Combining the contribution from all users followed by $i$ gives 
\begin{align}
    \frac{d\theta_i}{dt} & = \sum_{j\in V}\lambda_{ji}f(\theta_j-\theta_i)~~ \forall i\in V \label{eq:diff_eq_model}.
\end{align}
This set of differential equations will govern the opinion dynamics of users in a social network in our analysis.

We must select the opinion shift function $f$ for our model. The classic DeGroot model has $f(x)=\omega x$ where $\omega>0$ is the persuasion strength  \citep{degroot1974reaching}.   The DeGroot shift function is attractive, meaning the follower's opinion will be pulled towards the opinions of those they follow, no matter what the opinions are.  If one thinks about the implications of this with respect to very extreme opinions, one can begin to see some limitations of the model.  For instance, it is unlikely one would be influenced by an opinion vastly different from one's own, especially in highly polarized contexts such as political discussions.  This has been observed in different field experiments \citep{nyhan_when_2010, bail2018exposure, galasso_positive_2023} and validated in real social network data \citep{kozitsin_formal_2022}.  Another limitation of the DeGroot model can be seen if we consider the long-run behavior of the model.  It can be shown for most networks that if opinions  evolve according to the DeGroot model, then eventually they reach consensus \citep{acemoglu2011opinion,  hunter2022optimizing}.  However, research has shown that in actual social networks, consensus is not achieved. Rather, people's opinions exhibit persistent polarization \citep{adamic2005political, conover2011political, bakshy2015exposure, garimella2018political,rossetti2023bots}.    

In order for DeGoot's model to reflect these empirical findings, it needs to be modified in a way that somehow limits the persuasion strength.  One approach is to declare certain users to be stubborn agents, meaning their opinions do not change.  DeGroot's model under stubborn agents results in persistent polarization \citep{acemouglu2013opinion, ghaderi2013opinion,  vassio2014message, hunter2022optimizing}.  However, it is not clear how to determine which users are stubborn, though heuristics have been proposed \citep{des2022detecting}.  The stubborn agents model also allows for persuasion by extreme opinions, a phenomenon that does not reflect empirical findings.  There is a more natural modification of the DeGroot model known as the \emph{bounded confidence} model \citep{deffuant2000mixing, hegselmann_opinion_2002}.  In this model, a confidence interval $\epsilon$ is chosen and the shift function is given by 
\begin{align}
 f(x) = 
\begin{cases} 
\omega x & \text{if } |x| \leq \epsilon, \\
0 & \text{otherwise}.
\end{cases}
\label{eq:shift}
\end{align}
The shift function is non-zero only if the magnitude of the difference in the users' opinions is less than the confidence interval. This means that one cannot be persuaded by any opinion that deviates by more than $\epsilon$ from their own. Because of this, there are network configurations where the model exhibits persistent polarization \citep{lorenz2006consensus, blondel2009krause}. This bounded confidence model better captures phenomena observed in real social networks.  In this work  we will use it to model the opinion dynamics.


\section{Influence Campaign Problem}\label{sec:control}
We now formulate the problem of learning the actions of agents in an influence campaign on a social network.  We are given a social network $G= (V,E)$, initial opinion vector $\boldsymbol{\theta}(0) = \curly{\theta_i(0)}_{i\in V}$, posting rates $\curly{\lambda_{ij}}_{(i,j)\in E}$, and an opinion dynamics model in the form of equation \eqref{eq:diff_eq_model}.  The influence campaign will run from time $t=0$ to $t=T$.  We are also given an opinion objective function  $r(\boldsymbol{\theta}(T))$ that we want the agents to maximize.  We consider objectives that only depend on the value of the opinions at the end of the influence campaign.  

Our approach to influence campaigns accommodates various opinion objective functions. In this work, we focus on three specific objectives: maximizing the mean, minimizing the variance, and maximizing the variance. These objectives are chosen because they represent natural goals for influence campaigns.  Maximizing the mean opinion in the network is a common goal, as it aligns with the aim of gaining support for a particular topic. Conversely, minimizing the mean is equivalent to maximizing it if we invert the opinions' signs. Minimizing opinion variance aims to reduce polarization within the network, fostering greater consensus, which can lead to political stability and diminish the influence of extreme voices. In contrast, maximizing variance would increase polarization, potentially destabilizing a population—an objective that might be pursued by a nation seeking to weaken a geopolitical adversary.

The agents' policy in an influence campaign can be divided into two phases. First, the \textit{targeting policy} identifies targets within the network for the agents. Targeting users can be done in practice by simply following the targets and liking their content.  It has been found this is an effective technique to make a target follow an agent \citep{mosleh2021shared, yang2022mitigating}.  After selecting their targets, the agents' \emph{content policy} prescribes what content to post. In particular, this policy determines the opinion expressed in the agents' content.  The content policy is dynamic, which allows for the content opinion to evolve over time. 
The targeting and content policies determine the actions of the agents during the influence campaign.  The targeting policy is static and evaluated one time before the campaign begins.  After acquiring targets, the agents then implement the dynamic content policy by posting content with the appropriate opinion.  

To formally present the influence campaign problem, we now provide some convenient definitions.  The agents form a set $A$ and their targets are described by a binary $|A|\times|V|$ matrix $\mathbf{x}$, where $x_{ai}=1$ if user $i$ is a target of agent $a$ and zero otherwise.  We limit each agent to have a maximum of $d_{max}$ targets.  Having too many targets can make the agents appear to be spam or bots, which could result in them not being followed or getting blocked by their targets.  The opinion of content posted by agent $a$ at time $t$ is denoted as $u_a(t)$, and the opinions of all agents' at all times is denoted as $\mathbf{u}$.  We bound the opinions within a range $[u_{min}, u_{max}]$ which represent the most extreme opinions one can have on the given topic.  Agent $a$ posts content at a constant rate $\lambda_a$.  As with having too many targets, posting too rapidly can make the agents appear to be spam or bots.  Therefore, the agents' posting rate is capped at a value $\lambda_{max}$ which can be determined based on the activity levels of users in the specific social network. In our formulation we set $\lambda_a = \lambda_{max}$ for all agents, so the agents all post at the maximum feasible rate, as was done in \citep{hunter2022optimizing}.  This lets the agents exert maximum persuasion power while not appearing to be bots. 

The agents' policy for the influence campaign is determined by solving the following optimization problem:
\begin{align}
    \max_{\boldsymbol{x}, \boldsymbol{u}} ~~~&r(\boldsymbol{\theta}(T)) \nonumber \\
    \text{subject to} ~~~\frac{d\theta_i}{dt}=&\sum_{j\in V}\lambda_{ji}f(\theta_j(t)-\theta_i(t)) + \nonumber \\
    &\lambda_{max}\sum_{a\in A}x_{ai}f(u_a(t)-\theta_i(t)), ~~\forall i \in V,~ t\in [0,T], \nonumber \\
    ~~~&\sum_{i\in V}x_{ai}\leq d_{max}, ~~\forall ~a\in A, \nonumber \\
    ~~~&x_{ai} \in \{0,1\}, ~~\forall a\in A, ~i\in V, \nonumber \\
    ~~~&u_{a}(t) \in [u_{min}, u_{max}], ~~\forall ~a\in A,~ t\in [0,T]. \label{eq:optimal_control}
\end{align}
This optimization problem presents notable challenges. If the targets were fixed, determining the agents' opinions would be a nonlinear control problem, which is quite challenging to solve. There are methods such as collocation that can solve this control problem \citep{betts1998survey}. This method involves selecting specific points within the domain of the differential equations, termed collocation points, and ensuring that the differential equations are satisfied precisely at these points. The number of collocation points increases with the number of nodes in the network, which restricts the use of collocation to smaller networks. The selection of targets represents a discrete optimization problem, also scaling with the network's size. Solving for the targets and opinions jointly via collocation is feasible in principle, but only practical for small networks.

For larger networks with thousands of users, collocation methods become impractical. To address this, we develop a set of approximation algorithms that enable us to design influence campaign policies for large-scale networks. Our approach consists of two phases. First, we sequentially determine the target nodes for each agent in \(A\). For each agent \(a\), we solve for its target selection policy \(x_{ai}\), for all \(i \in V\). This process is repeated for all agents in \(A\). In the second phase, given the target nodes for all agents, we solve for their content policy \(u_a(t)\), for all $a\in A$ and \(t \in [0, T]\). The following sections provide a detailed explanation of this two-phase methodology.


\subsection{Content Policy}

We begin by assuming that the targets for all agents have been selected, so we can treat $x_{ai}$ as a constant.  The next step is to solve for the content policy on the interval $[0,T]$. To do this, we take the standard approach in control theory and write down the Hamiltonian function for the problem \citep{evans2005introduction}, which is given by
\begin{align*}
    H(\boldsymbol{\theta},\mathbf p) & = \sum_{i\in V}\sum_{j\in V}p_i(t)\lambda_{ji}f(\theta_j(t)-\theta_i(t)) + \lambda_{max}\sum_{a\in A}\sum_{i\in V} p_i(t)x_{ai}f(u_a(t)-\theta_i(t)),
\end{align*}
where the adjoint variables $p_i(t), \forall i \in V$ satisfy Hamilton's equations
\begin{align*}
    \frac{dp_i}{dt} & = -\frac{\partial H}{\partial \theta_i},~~\forall i \in V,\\
     \frac{d\theta_i}{dt}& =\frac{\partial H}{\partial p_i},~~\forall i \in V,
\end{align*}
and have boundary conditions 
\begin{align*}
    p_i(T) & = \frac{\partial r}{\partial \theta_i(T)},~~\forall i \in V.
\end{align*}
To obtain agent opinions, we use the Pontryagin's maximum principle which states that the optimal agent opinions will maximize the Hamiltonian at each time.  We note that the agents' opinions appear in the Hamiltonian only in the second summation.  Also, each agent's opinion appears in a separate term in the summation.  Using this observation, we find that the optimal opinion for agent $a$ at time $t$ is
\begin{align}
    u_a^*(t) & = \arg\max_{u_a}\sum_{i\in V} p_i(t)x_{ai}f(u_a-\theta_i(t)),~~~\forall a \in A\label{eq:optimal_opinion}.
\end{align}
We see from this expression that the agent's opinion maximizes a weighted sum of the opinion shifts of its targets, with the weights given by the adjoint variables.  These weights tell the agent which target it should focus on shifting.  For instance, in the DeGroot's model, this expression becomes
\begin{align*}
    u_a^*(t) & = \arg\max_{u_a} u_a\sum_{i\in V} p_i(t)x_{ai}\omega -\sum_{i\in V} p_i(t)x_{ai}\omega\theta_i(t)\\
    & = \arg\max_{u_a} u_a \omega \sum_{i\in V} p_i(t)x_{ai},~~~\forall a \in A.
\end{align*}
The optimal agent opinion is one of the end-points of its feasible region ($u_{min}$ or $u_{max}$), depending on the sign of the sum of the adjoint variables of its targets.  These opinion values maximize the magnitude of the agent's shift function.  The linearity of the shift function  under the DeGroot model results in a fairly simple content policy which is essentially direct advocacy.  No complex nudging is required by the agents.

For the bounded confidence model, which has a non-linear shift function, such a simple result will not generally be possible.  To get a better understanding of what the agent's policy is under the bounded confidence model, we consider the simple case where agent $a$ has a single target which we assume is node 1. In this case, the agent's opinion which maximizes the Hamiltonian is
\begin{align}\label{agent_opinion_1-node}
    u_a^*(t) = \begin{cases} 
\theta_1(t) - \epsilon & \text{if } p_1(t) < 0, \\
\theta_1(t) + \epsilon & \text{otherwise},
\end{cases}
\end{align}
where $\epsilon$ is the confidence interval parameter from equation \eqref{eq:shift}.
This analysis reveals that the agent's opinion is always within an $\epsilon$-neighborhood of its target's opinion. If the agent's opinion falls outside the target's confidence interval, the shift function is zero, and no persuasion can occur. The shift function attains its maximum magnitude precisely at the edge of the interval. Consequently, the agent's opinion is always slightly above or below its target's opinion. We say that the agent is \emph{nudging} its target in the desired direction. The choice of direction for the nudge is determined by the sign of the adjoint variable.  When dealing with multiple targets, the agent's opinion will lie somewhere between the lowest target opinion minus $\epsilon$ and the highest target opinion plus $\epsilon$. The precise location is determined by the values of the adjoint variables, which as mentioned before, tell the agent which targets to shift.   

If we knew the adjoint variables, finding the agents' opinions is straightforward.  Unfortunately, to obtain the adjoint variables one must solve a non-linear system of coupled differential equations with split boundary conditions (the opinions are specified at $t=0$ but the adjoint variables are specified at $t=T$).  This problem is difficult to solve for the bounded confidence model because of the non-linear shift function.  However, our analysis here shows that the agent opinion maximizes the weighted sum of its targets' opinion shifts, with the targets weighted by the adjoint variables.  With that in mind, we now consider the following approximation that has been used in previous work
\citep{chen_shaping_2024} to the control problem in \eqref{eq:optimal_control}.  We rewrite the objective as
\begin{align*}
    \max r(\boldsymbol{\theta}(T)) & = \max [r(\boldsymbol{\theta}(0)) + \int_{0}^T\frac{dr}{dt}dt]\\
    & = \max \int_{0}^T\frac{dr}{dt}dt.
\end{align*}
The control problem involves maximizing this integral subject to the opinion dynamics constraints, which is difficult to do for large networks and with non-linear dynamics.  Instead, we will employ a greedy approach and sequentially maximize the integrand ${dr}/{dt}$ at each time $t$.  We will see that this allows us to solve for the agents' opinions even on large networks.  We expand the integrand, replacing ${d\theta_i}/{dt}$ with the opinion dynamics model in equation \eqref{eq:diff_eq_model}, to obtain
\begin{align*}
    \frac{dr}{dt} & = \sum_{i\in V}\frac{\partial r}{\partial \theta_i(t)}\frac{d\theta_i(t)}{dt}\\
   & =  \sum_{i,j\in V}\frac{\partial r}{\partial \theta_i(t)} \lambda_{ji}f(\theta_j(t)-\theta_i(t)) + \lambda_{max}\sum_{i\in V, a\in A}\frac{\partial r}
   {\partial \theta_i(t)}x_{ai}f(u_a(t)-\theta_i(t)).
\end{align*}

We see that the integrand includes the agents' opinions in separate additive terms in the second summation.    Thus, we can optimize over each agent's opinion separately. The opinion of each agent $a\in A$ at a time $t$ is given by
\begin{align}
    u^*_a(t) &= \arg\max_{u_a}\sum_{i\in V}\frac{\partial r}{\partial \theta_i(t)} x_{ai}f(u_a-\theta_i(t))\label{eq:greedy_opinion}.
\end{align}
Here we drop $\lambda_{max}$ from the expression since it is a constant. We see that the greedy agent opinion has a very similar form to the optimal agent opinion in equation \eqref{eq:optimal_opinion}, except that we have replaced the adjoint variables $p_i(t)$ with the partial derivative of the reward with respect to $\theta_i(t)$. This will lead the agent to concentrate on targets whose current opinions have the greatest impact on the objective.

We find that the optimization in equation \eqref{eq:greedy_opinion} often leads to the agent opinion having large jumps.  This is often due to the agent shifting its focus between targets with non-overlapping confidence intervals.  In practice, wildly varying opinions will make the agent appear non-human and may reduce its persuasive ability.  In order to avoid this, we put a constraint on how much the agent opinion can vary.  In practice, the agent opinions are calculated at discrete times, typically days, which we denote $t_0=0, t_1, t_2, ..., t_n = T$.  We limit the agent's opinion to change by no more than a regularization constant $\gamma$ between consecutive days.  Formally, when solving for the agent's opinion at time $t_{i+1}$ using equation \eqref{eq:greedy_opinion}, we constrain its value to be in the interval $[u^*_a(t_i)-\gamma, u^*_a(t_i)+\gamma]$ where $u^*_a(t_i)$ is the value of the opinion at the previous time step.  We find that setting $\gamma = 0.001$ is a reasonable limit on the daily change of the agent opinion given results on persuasion in the literature \citep{pink2021elite, pink2023effects, bai2023artificial}.

In a multi-agent system, each agent's content policy is determined concurrently at every time step. Initially, each agent is assigned a unique set of targets, ensuring that the target sets do not overlap between agents. At each time step, every agent updates its opinion simultaneously by solving equation~\eqref{eq:greedy_opinion}. The opinions of all nodes then evolve according to equation $\eqref{eq:diff_eq_model}$ under the bounded confidence model until the next update cycle, when agents recompute their opinions. This procedure allows each agent to account for both its assigned targets and the content history of other agents during the update process. The overall algorithm is summarized in Algorithm~\ref{algo:opinion}. In Section~\ref{sec:targeting}, we detail how the target matrix $\mathbf{x}$ is obtained.

\begin{algorithm}
\caption{Content Policy Algorithm}
\label{algo:opinion}
\begin{algorithmic}[0]
\State \textbf{Input:} $G=(V,E,\lambda)$, $\boldsymbol{\theta}(0)$, $T$, $r(\cdot)$, $A$, $\lambda_{max}$, $\mathbf{x}$
\For{$t = 0,1,\dots,T-1$}
    \State $u^*_a(t) \gets \arg\max_{u_a}\sum_{i\in V}\frac{\partial r}{\partial \theta_i(t)} x_{ai}f(u_a-\theta_i(t))$ ~~$\forall a \in A$\\
    \State $\theta_i(t+1) \gets \theta_i(t) + \displaystyle\int_{t}^{t+1} \left[\sum_{j\in V}\lambda_{ji}f\big(\theta_j(t)-\theta_i(t)\big) + \lambda_{\max}\sum_{a\in A}x_{ai}f\big(u_a^*(t)-\theta_i(t)\big)\right]\, dt\quad \forall i \in V$
\EndFor
\State \textbf{Output:} $\boldsymbol{u}$, $\boldsymbol{\theta}(T)$
\end{algorithmic}
\end{algorithm}

We can gain intuition for the greedy agent opinions by studying the partial derivatives of the opinion objectives in more detail.  Table \ref{table:deriv} lists these partial derivatives of the opinion objectives that we consider here.  We first consider maximizing the opinion mean.  For this objective we see that the partial derivative is a positive constant.  This means that the agent tries to pull up the opinion of its target, but it does not prefer any specific target.  If the agent wants to minimize the opinion variance, then it will either try to pull up nodes whose opinions are below the mean or pull down nodes whose opinions are above the mean, which would have the effect of reducing the variance.  The opposite occurs if the objective is maximizing the variance.  


\begin{table}[!hbt]
\centering
\renewcommand{\arraystretch}{1.75} 
\caption{Table of the partial derivatives for different objective functions.  We have used $\mu$ to refer to the mean of the opinions.}
\label{table:deriv}
\begin{tabular}{|c|c|c|}
\hline
$\textbf{Objective}$ & $\textbf{Objective Function}$ r & $\textbf{Partial Derivative}$ $\frac{\partial r}{\partial\theta_i}$ \\[0.2cm] \hline
 Maximize Mean &     $\dfrac{1}{|V|}\sum_{i\in V}\theta_i$ & $\dfrac{1}{|V|}$ \\[0.2cm] \hline
Minimize Variance & $-\dfrac{1}{|V|-1}\sum_{i\in V}(\theta_i-\mu)^2$ & $-\dfrac{2}{|V|-1}(\theta_i-\mu)$ \\[0.2cm] \hline
Maximize Variance & $\dfrac{1}{|V|-1}\sum_{i\in V}(\theta_i-\mu)^2$ & $\dfrac{2}{|V|-1}(\theta_i-\mu)$ \\[0.2cm] \hline
\end{tabular}
\end{table}

\subsection{Targeting Policy}\label{sec:targeting}

We now turn to the selection of targets for the agents. This problem was originally investigated for opinion diffusion models \citep{kempe2003maximizing, kempe_influential_2005} and has more recently been studied in the context of the DeGroot model  \citep{vassio2014message, hunter2022optimizing}. In both scenarios, a greedy approach was utilized, where targets were chosen sequentially based on the marginal increase in the opinion objective they provided. For both the opinion diffusion and DeGroot models, it was demonstrated that objectives such as number of users reached or opinion mean are monotone submodular functions \citep{kempe2003maximizing, hunter2022optimizing}, allowing performance bounds to be established for greedy maximization \citep{nemhauser1978analysis}. However, the opinion variance under the DeGroot model was shown to be neither monotone nor submodular \citep{hunter2022optimizing}. Despite this, a greedy approach was found to be effective for both minimizing and maximizing variance on real Twitter networks. In this work, we will adopt a similar approach, but with necessary modifications to the greedy targeting under the bounded confidence model.

The greedy approach with the DeGroot model focused on shifting the equilibrium opinions, for which a closed form expression exists.  Unfortunately, no such expression exists for the equilibrium opinions in the bounded confidence model due to the non-linearity of its shift function.  However, in our formulation of the influence campaign problem, we are concerned with the opinions at a final time $T$, so it is more natural to select targets based on the opinions at this time rather than at equilibrium.  We can obtain the opinions at any finite time by numerically integrating the opinion dynamics in equation \eqref{eq:diff_eq_model}.  To do this, we need to know the agents' content policy, which we can obtain using equation \eqref{eq:greedy_opinion}.  This forms the basis of our greedy targeting policy, which we now present. 

The targeting algorithm is given the objective function $r(\cdot)$, network $G=(V,E, \lambda)$, initial opinions $\boldsymbol{\theta}(0)$, a set $A$ of agents and a specified number $d_{max}$ of targets for each agent.  We initialize the target assignment matrix $\mathbf x=\mathbf 0$ and the current best objective value $r^*=0$ .   We choose the targets in a greedy manner, iterating over the agents (the agent iteration), and then for each agent iterating over the potential targets (the target iteration).  To limit the runtime of the targeting algorithm, we will not consider all nodes in the network as potential targets.   Instead, we will only select targets from a consideration set $C\subset V$, as was done in \citep{hunter2022optimizing}.  A simple way to construct this set is to consider the nodes with a high network centrality score.  There are a variety of centralities to choose from, but we find that out-degree works well, as this gives high scores to nodes with a large direct reach in the network.  In addition to $C$, we also define the set of selected targets as $S$ to keep track of which users have been assigned to an agent.  This is to avoid multiple agents having the same target.

At each targeting iteration, agent $a$ selects a candidate node $i$ from the available set $C \setminus S$ and temporarily sets $x_{ai}=1$. With the updated target matrix $\mathbf{x}$, Algorithm~\ref{algo:opinion} is executed: all agents concurrently compute their content policies $u^*_a(t)$ via equation~\eqref{eq:greedy_opinion} at every time step, and the opinions of all nodes evolve under the bounded confidence model according to equation~\eqref{eq:diff_eq_model} until the next update cycle, at which point the agents recompute their opinions. This process continues until the final time $T$ is reached. For brevity, let $r(\mathbf{x})$ denote the objective value obtained from this process with $x_{ai}=1$. If $r(\mathbf{x}) > r^*$, node $i$ is permanently added to agent $a$'s target set; otherwise, $x_{ai}$ is reset to 0. Agent $a$ repeats this targeting process until either all available nodes in $C$ have been evaluated or the number of targets reaches the maximum limit $d_{\max}$. This procedure is then applied to every agent in $A$. The pseudo-code for this targeting policy is summarized in Algorithm~\ref{algo:target}.

\begin{algorithm}
\caption{Targeting Policy Algorithm}
\label{algo:target}
\begin{algorithmic}[0]
\State \textbf{Input:} $G=(V,E,\lambda)$, $\boldsymbol{\theta}(0)$, $T$, $r(\cdot)$, $A$, $\lambda_{max}$, $d_{\max}$, $C$
\State $S \gets \emptyset$, $\mathbf{x} \gets \mathbf{0}$, $r^* \gets 0$
\For{$a \in A$}
    \State $d_a \gets 0$
    \For{$i \in C \setminus S$}
        \State $x_{ai} \gets 1$
        \State Compute final opinions $\boldsymbol{\theta}(T)$ using Algorithm~\ref{algo:opinion}
        \State $r(\mathbf{x}) \gets r(\boldsymbol{\theta}(T))$
        \If{$r(\mathbf{x}) > r^*$}
            \State $S \gets S \cup \{i\}$, $d_a \gets d_a + 1$, $r^* \gets r(\mathbf{x})$
        \Else
            \State $x_{ai} \gets 0$
        \EndIf
        \If{$d_a \geq d_{\max}$}
            \State \textbf{break}
        \EndIf
    \EndFor
\EndFor
\State \textbf{Output:} $\mathbf{x}$
\end{algorithmic}
\end{algorithm}

\subsection{Performance Guarantees}
Our targeting and content policies both utilize a greedy heuristic. We aim to establish performance guarantees for these algorithms across different opinion objectives. Such results have already been demonstrated for the DeGroot model, where the opinion objective is evaluated at the equilibrium opinions and treated as a set function of the targets. It was shown that the opinion mean is monotone and submodular, while the opinion variance is not \citep{hunter2022optimizing}. In this work, we establish similar results for the bounded confidence model. The proofs for these findings can be found in \ref{sec:proof_theorems}.

Under the DeGroot model the opinion mean was shown to be submodular.  We find that under the bounded confidence model this ceases to be true.  We have the following result.
\begin{theorem}[Non-Submodular Targeting]
    Consider the bounded confidence model of opinion dynamics at any finite time $T$. Suppose a fixed-opinion agent targets a subset $S$ of nodes, thereby influencing their opinions. Let $r(S)$ denote an objective function evaluated on the opinions at time $T$, where $r(S)$ is either the opinion mean or the opinion variance. Then, the function $r(S)$ is not submodular with respect to the targeted nodes.
    \label{thm:not_submodular_under_BC}
\end{theorem}

This result shows that submodularity does not even hold for the opinion mean under the bounded confidence model.  In other words, the diminishing returns property does not hold for these objectives under targeting in the bounded confidence model. Therefore, there is no simple performance guarantee we can provide for the greedy approach. 
However, we will find that this approach will still be effective in increasing the opinion objective on real networks.

We next consider the impact of increasing the number of agents utilized in the influence campaign.  Intuitively, we would expect more agents to be beneficial as each agent can focus shifting users in different segments of the opinion spectrum under the bounded confidence model.   We find that this is the case if one can calculate the optimal opinion policy.  Formally, we have the following result.

\begin{theorem}[Optimality of Adding Agents]
    Let $r_k^*$ be the optimal objective value of the influence campaign problem in \eqref{eq:optimal_control} with $k$ agents.  Then we have for any integer $k\geq 0$
    $$r_{k+1}^* \geq r_k^*.$$
    \label{thm:adding_agents_optimal}
\end{theorem}
This result arises because each new agent introduces additional targets, providing extra degrees of freedom to optimize the opinion objective. Since the new agent can choose not to target anyone, the optimal objective value cannot decrease. However, by optimizing the new agent's targets, it is possible to further improve the objective.

Key to Theorem \ref{thm:adding_agents_optimal} is that one is able to compute the optimal opinion policy.  However, we have found that this is computationally challenging for large networks.  Instead, we utilize the greedy content policy in \eqref{eq:greedy_opinion}.  It turns out that under this approach, additional agents can sometimes decrease the opinion objective.  We articulate this observation in the following proposition.

\begin{proposition}[Limitations of Adding Agents Under a Suboptimal Policy]
Let $r_k$ be the objective value of the influence campaign problem in \eqref{eq:optimal_control} with $k$ agents, using the greedy content and targeting policies in Algorithms \ref{algo:opinion} and \ref{algo:target}. There exist network configurations and integers $k$ and $k'$ such that $k < k'$ but \begin{align*} r_k > r_{k'}. \end{align*} \label{prop:adding_agents_suboptimal}
\end{proposition}
This implies that adding more agents under a suboptimal policy can, counterintuitively, reduce the objective value. Consequently, without optimal content and targeting strategies, merely increasing the number of agents does not ensure better outcomes. In some cases, adding agents may even worsen performance. Therefore, it is essential to perform numerical simulations with varying agent numbers to identify the most effective configuration in practice.

\section{Results}\label{sec:results}
We now present simulation results for our influence campaign policies on different network topologies.  We begin with synthetic networks to provide insights to our algorithms. We then scale our policy to large-scale Twitter networks.  We compare our nudging-based policies to DeGroot-based policies in a single-agent setting, and further demonstrate that employing multiple agents leads to improved opinion objectives. Throughout this section, we limit the agent opinion and all nodes' opinions to be between zero and one.

\subsection{Synthetic Networks}
We employ two synthetic networks to better understand the content and targeting policies. To analyze content policies, we use a simple two-node path network. For the targeting policy, we utilize a slightly larger ten-node path network. Despite their simplicity, these structures provide valuable intuition into the policies. Both synthetic networks have bounded confidence model parameters $\epsilon = 0.1$ and $\omega = 0.003$.

\subsubsection{Two-Node Network}
\begin{figure}[h]
    \centering
    \includegraphics[width=\textwidth]{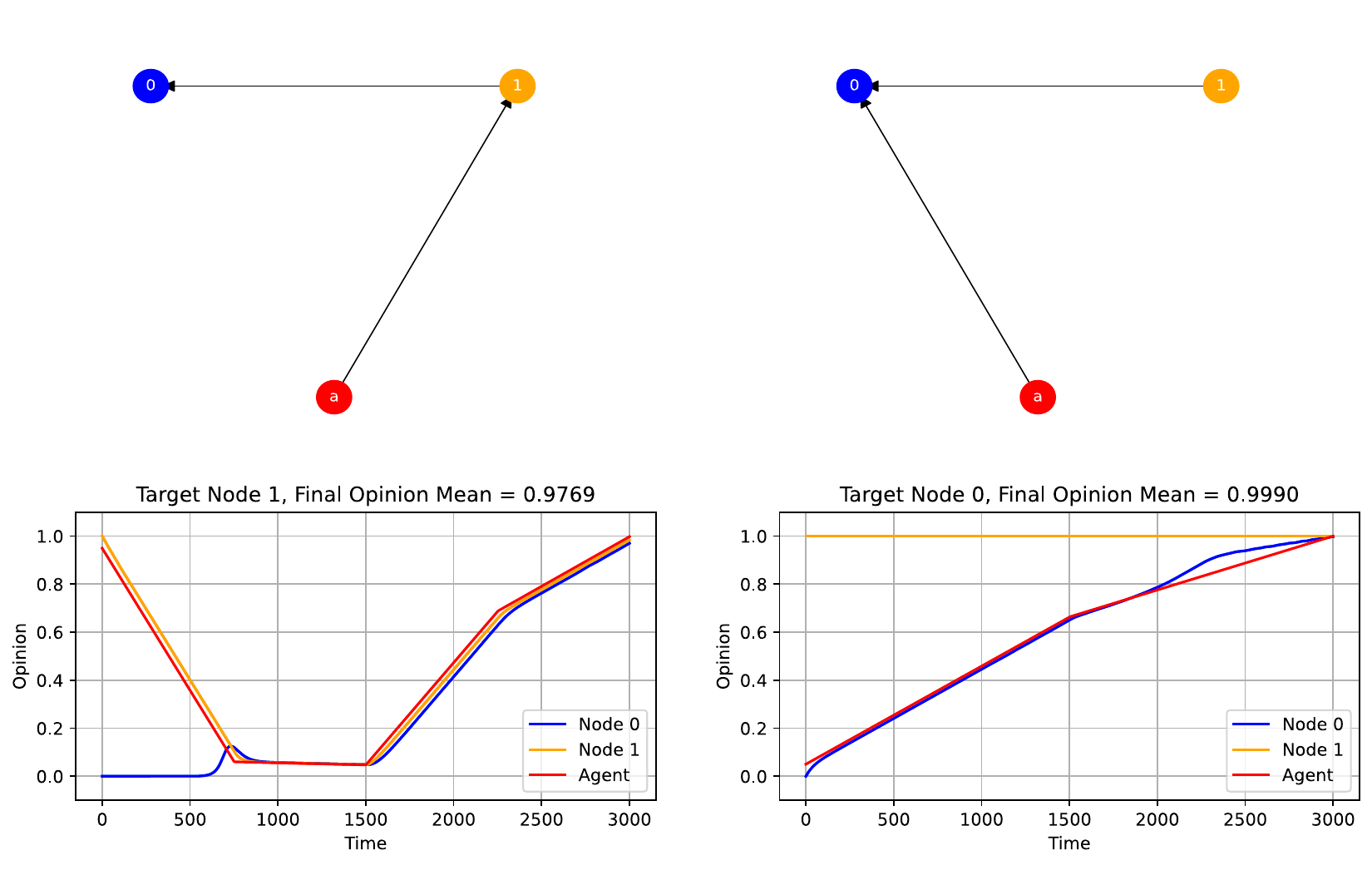}
    \caption{\textbf{Top:} The two-node network with the agent targeting node 1 (left) or node 0 (right).\\
    \textbf{Bottom:} The opinions of agent and nodes, obtained by the collocation method for the two-node network with agent targeting node 1 (left) or node 0 (right). The objective is to maximize the opinion mean.} 
    \label{fig:2_node_evolution}
\end{figure}

We begin with a simple two-node network shown in Figure \ref{fig:2_node_evolution}.  In this network, node 1, which has an opinion of one, is followed by node 0, which has an opinion of zero. We assume that a single agent can target only one of the nodes. The nodes and the agent all tweet at a rate of ten tweets per unit time. For each target choice, Figure \ref{fig:2_node_evolution} shows the resulting agent opinions obtained by the collocation method. When the target is node 1, the agent ends up initially lowering its opinion to bring the opinion of node 1 down to that of node 0, followed by an increase in its opinion to bring up both nodes' opinions.  This exemplifies a bang-bang type persuasion strategy. While this policy is mathematically optimal, such extreme opinion shifts are unrealistic. It is unlikely that a human would drastically change their stance only to return to their original opinion. Such rapid opinion changes from the agent can make it appear unnatural. A more effective and realistic approach is to target node 0. If the agent targets node 0, as seen in Figure \ref{fig:2_node_evolution} the collocation solution has the agent gradually nudge the target upward, without any oscillating opinion.

\begin{figure}[h]
    \centering
    \includegraphics[width=\textwidth]{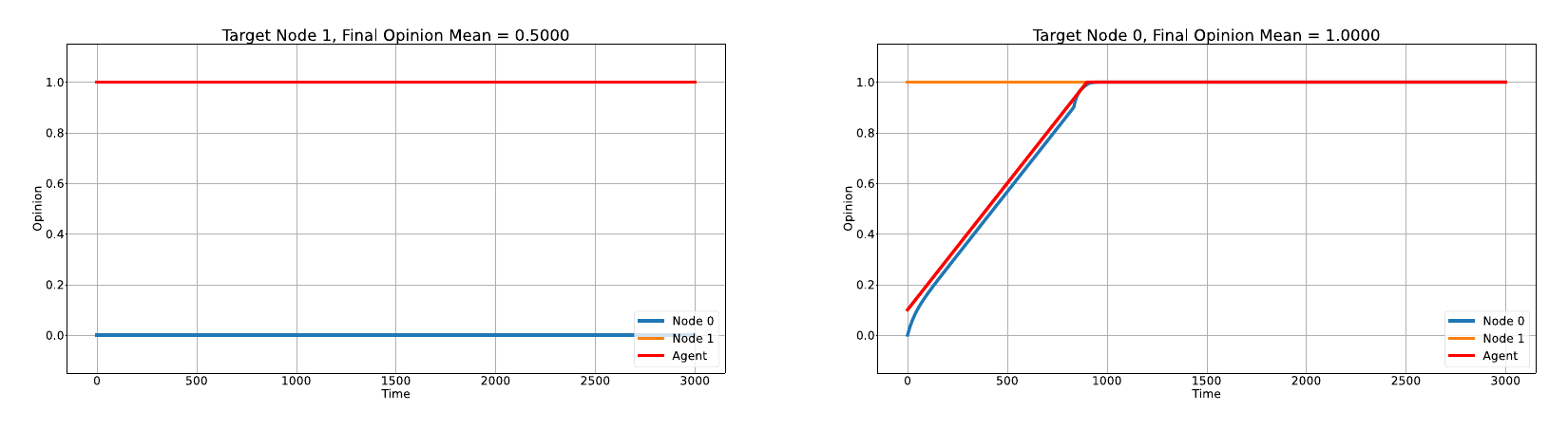}

    \caption{Agent opinions under the greedy content policy for a two-node network, where the agent targets node 1 (left) or node 0 (right) to maximize the mean opinion. Under our greedy targeting policy, the agent will select node 0 as its target. }  
    \label{fig:2_node_evolution_NG}
\end{figure}

We next consider the types of policies obtained for the two-node network when using our greedy nudging content policies given a set of targets.  We see in Figure \ref{fig:2_node_evolution_NG} that if node 0 is targeted, our nudging policy gradually increases the agent opinion over time. However, if the target is node 1,  the nudging policy does not persuade node 0 at all.  This highlights that the content policy alone is insufficient; successful persuasion requires intelligent targeting and  content policies. 

When applying greedy targeting to this network, it turns out that the agent
selects node 0, and the resulting nudging policy is able to achieve the maximum objective.  Therefore, with greedy targeting the agent can focus on nodes where a nudging policy is effective.  This example highlights how our nudging content policy can not only make agents more effective at persuasion, but also appear more human-like.

\subsubsection{Path Network}
Next, we create a ten-node path network to show how greedy targeting performs audience segmentation. We will see that the policy ends up having each agent target users whose opinions fall within a range dictated by the size of the confidence interval.  This way the agents are able to persuade all of their targets simultaneously. In our simulation, the ten nodes have initial opinions evenly spaced between zero and one, and each node tweets at a rate of one tweet per unit time. We use three agents, each tweeting at a rate of ten tweets per unit time and having a targeting budget of two followers.

\begin{figure}[h]
    \centering
    \includegraphics[width=\textwidth]{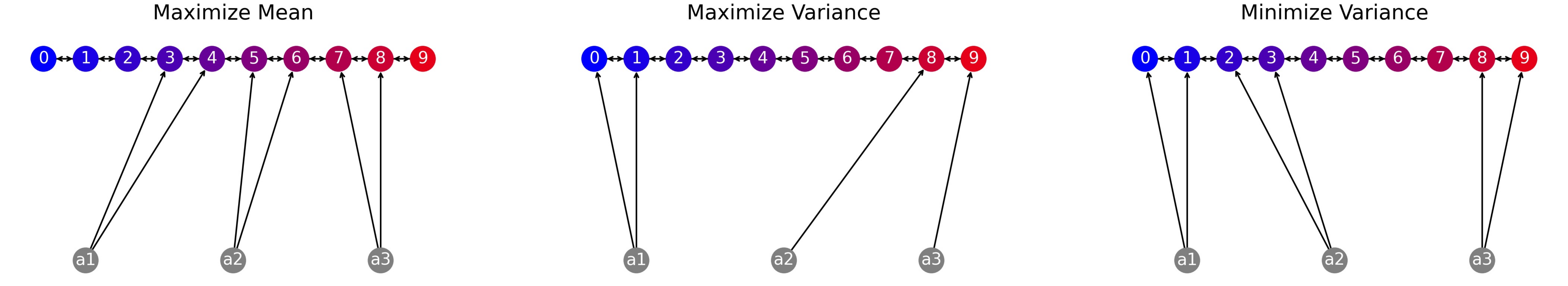}
    \caption{Targets of the three agents to persuade a ten-node path network, for the objectives of maximizing the mean (left), maximizing the variance (middle), and minimizing the variance (right).}
    \label{fig:targets_3_agents_test}
\end{figure}

\begin{figure}[h]
    \centering
    \includegraphics[width=\textwidth]{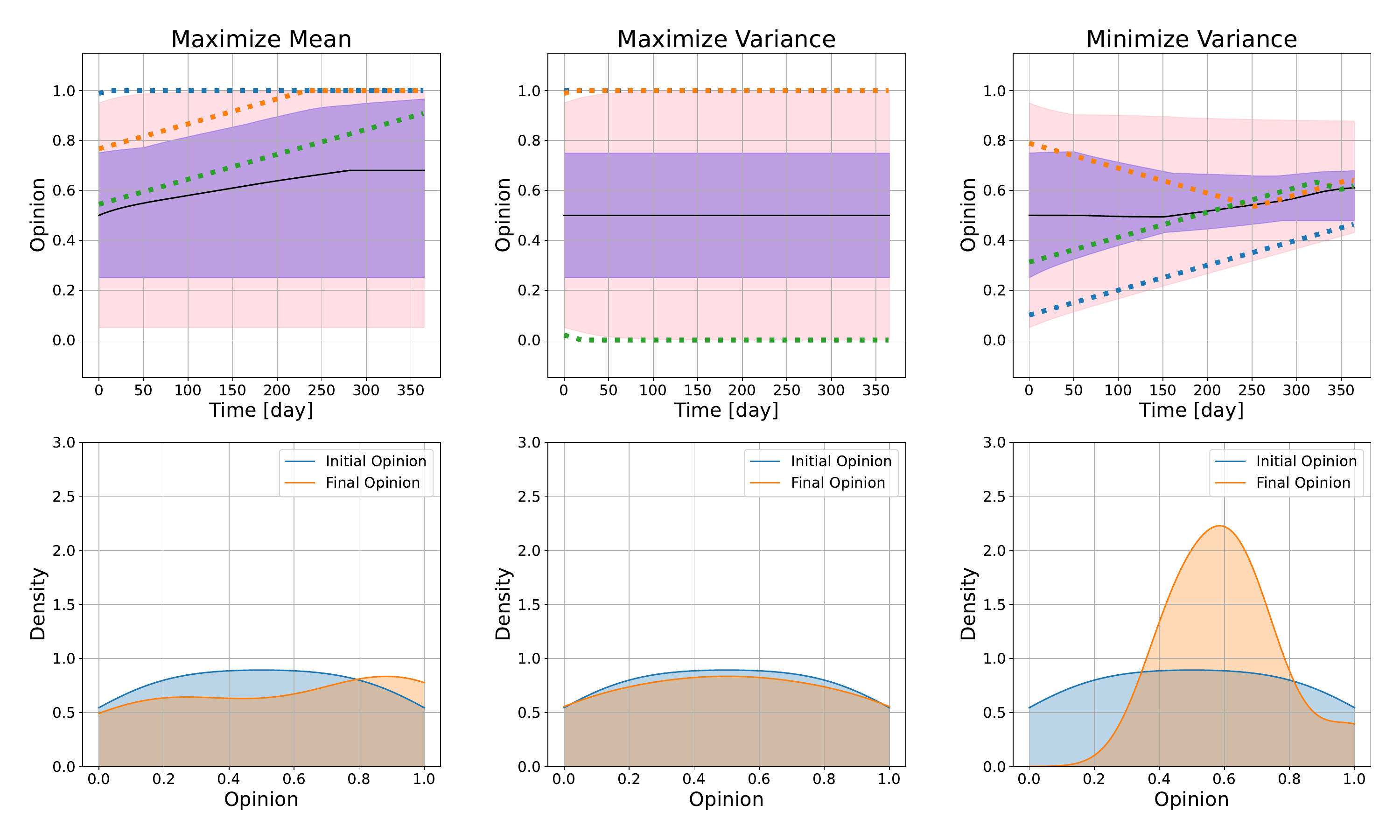}
    \caption{Opinion time series (top row) and density plots (bottom row) for different objectives under three nudging agents on a ten-node path network. The time series depicts opinion distributions (shaded regions) and agent opinions (dashed lines). The purple region represents the 25th to 75th quantiles, while the pink region represents the 5th to 95th quantiles. Opinion densities are plotted for initial (blue) and final (orange) opinions. The objectives are to maximize mean (left), maximize variance (middle), and minimize variance (right).}
    \label{fig:nudging_3_agents_test}
\end{figure}

As shown in Figure \ref{fig:targets_3_agents_test}, for each objective, greedy targeting assigns each agent to a different segment of the opinion spectrum, effectively performing audience segmentation. This segmented targeting, combined with our agent nudging policy, successfully shifts opinions in the desired direction, as demonstrated in Figure \ref{fig:nudging_3_agents_test}. For the objective of maximizing the mean, the agents focus on moving the upper opinion quantiles further upward, resulting in a mode around 0.9 in the final opinion density plot. For maximizing variance, the agents concentrate on pulling the extreme nodes on both ends further apart. For minimizing variance, the agents start by targeting the extreme opinions and gradually pull them toward an opinion around 0.6, effectively bringing the opinion closer together. In the density plot, this results in a mode centered around 0.55, indicating reduced polarization. This example illustrates how our targeting policy effectively performs audience segmentation, with multiple agents employing a divide-and-conquer approach to persuade the network.

\subsection{Twitter Networks}
\subsubsection{Datasets}
This subsection compares the effectiveness of nudging agents across two distinct Twitter network topologies, each associated with significant social events: the 2016 U.S. presidential election and the 2019 Gilets Jaunes protests in France. The key properties of these networks are summarized in Table \ref{table:network}. Following the simulation settings in \citep{chen_shaping_2024}, we randomly sample a subgraph of 30,000 nodes from each network and set the bounded confidence model parameters to $\epsilon = 0.1$ and $\omega = 0.003$. The agent tweet rate $\lambda_{max}$ is ten per day as it is the maximum tweet rate observed in the U.S. election dataset. We assume a total influence budget of 100 followers, representing 0.3\% of the nodes in the network. Due to computational limitations in the greedy search for targets, we restrict our candidate pool to the top 1,000 nodes with the highest out-degree (follower count) in the network. From this pool, 100 followers are selected. The agent's content policy regularization parameter, $\gamma$, is set to 0.001 to ensure that the resulting nudging content policy does not change rapidly.

\begin{table}[!hbt]
\centering
\caption{Basic information about the Twitter datasets. M is millions and K is thousands.}
\label{table:network}
\begin{tabular}{@{}p{3cm}p{3.5cm}p{2.5cm}p{2.5cm}p{2.5cm}@{}}
\toprule
\textbf{Event} & \textbf{Data Collection Period} & \textbf{Number of Tweets} & \textbf{Number of Follower Edges} & \textbf{Number of Users} \\ \midrule
U.S. Presidential Election & Jan. 2016--Nov. 2016 & 2.4M & 5.4M & 78K \\ \midrule
Gilets Jaunes            & Jan. 2019--Apr. 2019 & 2.3M & 4.6M & 40K \\ \bottomrule
\end{tabular}
\end{table}

\subsubsection{DeGroot Policy versus Nudging Policy}
Our analysis begins by comparing the persuasion capabilities of two influence campaign policies within these networks: the DeGroot policy and the nudging policy. For each policy, we employ one agent and evaluate the final objectives at $T=365$ under the bounded confidence opinion dynamics model.

Under the DeGroot policy, the agent maintains a fixed opinion over time. This approach simplifies policy computation but does not account for the bounded confidence effect. When the goal is to maximize the mean, the agent holds a constant opinion of 1; for variance maximization, the agent holds a constant opinion of 0 (chosen arbitrarily between 0 and 1); and for variance minimization, the agent maintains an opinion of 0.5. For target selection, the DeGroot policy assumes that all nodes update their opinions according to linear dynamics with the parameter value $\omega=0.003$, and targets are chosen by simulating these dynamics from  $t=0$ to a finite time $T=365$. Since node opinions evolve gradually, they do not reach equilibrium within this simulation horizon.

In contrast, our nudging policy enables the agent to adjust its opinion dynamically over time using Algorithm~\ref{algo:opinion}, thereby incorporating the bounded confidence effect. This policy assumes that all nodes update their opinions according to the bounded confidence model and selects targets using Algorithm~\ref{algo:target}. To reduce computational complexity associated with the nonlinearity of the bounded confidence model, we simulate opinion updates from $t=0$ to $T=30$ during the targeting phase, even though the final objectives when comparing to the DeGroot policy are evaluated at $T=365$.

Table~\ref{table:policy_assumptions} summarizes the content and targeting assumptions for both policies.

\begin{table}[!hbt]
\centering
\caption{Influence Campaign Policies and Their Assumptions}
\label{table:policy_assumptions}
\begin{tabular}{@{}p{3cm}p{5cm}p{5cm}@{}}
\toprule
\textbf{Influence Campaign Policy} & \textbf{Content Assumption} & \textbf{Targeting Assumption} \\ \midrule
DeGroot Policy & Agent opinion is fixed over time. & All nodes follow linear opinion dynamics. Simulation period $T=365$. \\ \midrule
Nudging Policy & Agent opinion is adaptive over time. & All nodes follow bounded confidence opinion dynamics. Simulation period $T=30$. \\ \bottomrule
\end{tabular}
\end{table}

\begin{figure}[h]
    \centering
    \includegraphics[width=\textwidth]{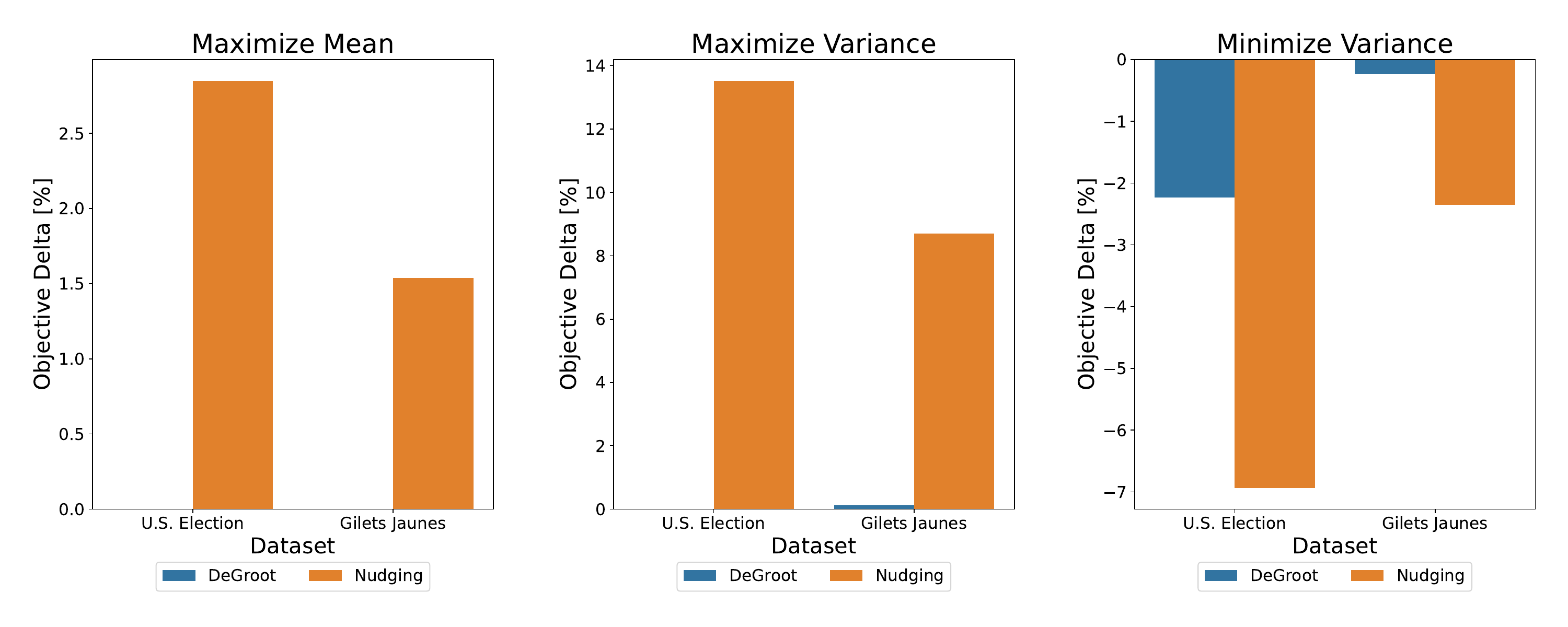}
    \caption{Bar plots comparing the change in final objective values for different Twitter network datasets relative to no agent  under DeGroot (blue) and nudging (orange) policies. The objectives are maximizing mean (left), maximizing variance (middle), and minimizing variance (right). The y-axis ''Objective Delta'' measures the percentage change relative to the objective without an agent. }
    \label{fig:objectives_DeGroot_nudging}
\end{figure}

Figure \ref{fig:objectives_DeGroot_nudging} shows the resulting final objectives for the different influence campaign policies.  We plot how much the objective under policies differs from the objective with no agent.  Nudging outperforms the DeGroot policy for all objectives.  The improvement percentages range from moderate to substantial, underscoring the effectiveness of adaptive contents over static contents when people exhibit the bounded confidence effect.  For instance, we see that with the DeGroot policy, there is usually no difference between the objective with and without a DeGroot agent.  This policy is not able to persuade because the agent's opinion is outside the confidence interval of its targets.

Moreover, we find that sometimes the two policies select different numbers of targets, as shown in Table \ref{table:number_targets}. We find that the DeGroot policy always fills its 100-node targeting budget.  This is because under the DeGroot model, an agent can persuade anyone, so having more targets can only improve performance.  However, the nudging policies sometimes choose many fewer targets than allowed by their budget. This is the case with the Gilets Jaunes dataset, where the nudging policy chooses almost an order of magnitude fewer targets than the DeGroot policy.  This difference stems from the nudging model's incorporation of bounded confidence, where influence is limited to individuals whose opinions fall within a certain confidence interval. As a result, sometimes adding another target will not improve the objective at all. This phenomenon is more pronounced in the Gilets Jaunes network than in the U.S. election network, possibly due to differences in tweeting frequency among users. Although the agent tweets at a rate of ten per day in both networks, the Gilets Jaunes network exhibits a higher median tweet rate (0.13 tweets per day) compared to the lower median tweet rate (0.02 tweets per day) in the U.S. election network. The higher tweeting activity among nodes in the Gilets Jaunes network diminishes the marginal impact of the agent's persuasion efforts, leading to fewer targets being selected.


\begin{table}[!hbt]
\centering
\caption{Number of targets selected by one agent for both the U.S. election and the Gilets Jaunes datasets. Each table shows results for DeGroot policy (top) and nudging policy (bottom) under objectives of maximizing the mean (left), maximizing variance (middle), and minimizing variance (right).}
\label{table:number_targets}
\begin{tabular}{lccc}
\toprule
& \textbf{Maximize Mean} & \textbf{Maximize Variance} & \textbf{Minimize Variance} \\
\midrule
\multicolumn{4}{l}{\textbf{U.S. Election}} \\
\midrule
DeGroot Policy & 100 & 100 & 100 \\
Nudging Policy & 100 & 100 & 100 \\
\midrule
\multicolumn{4}{l}{\textbf{Gilets Jaunes}} \\
\midrule
DeGroot Policy & 100 & 100 & 100 \\
Nudging Policy & 7 & 11 & 16 \\
\bottomrule
\end{tabular}
\end{table}

\begin{figure}[h]
    \centering
    \includegraphics[scale=0.4]{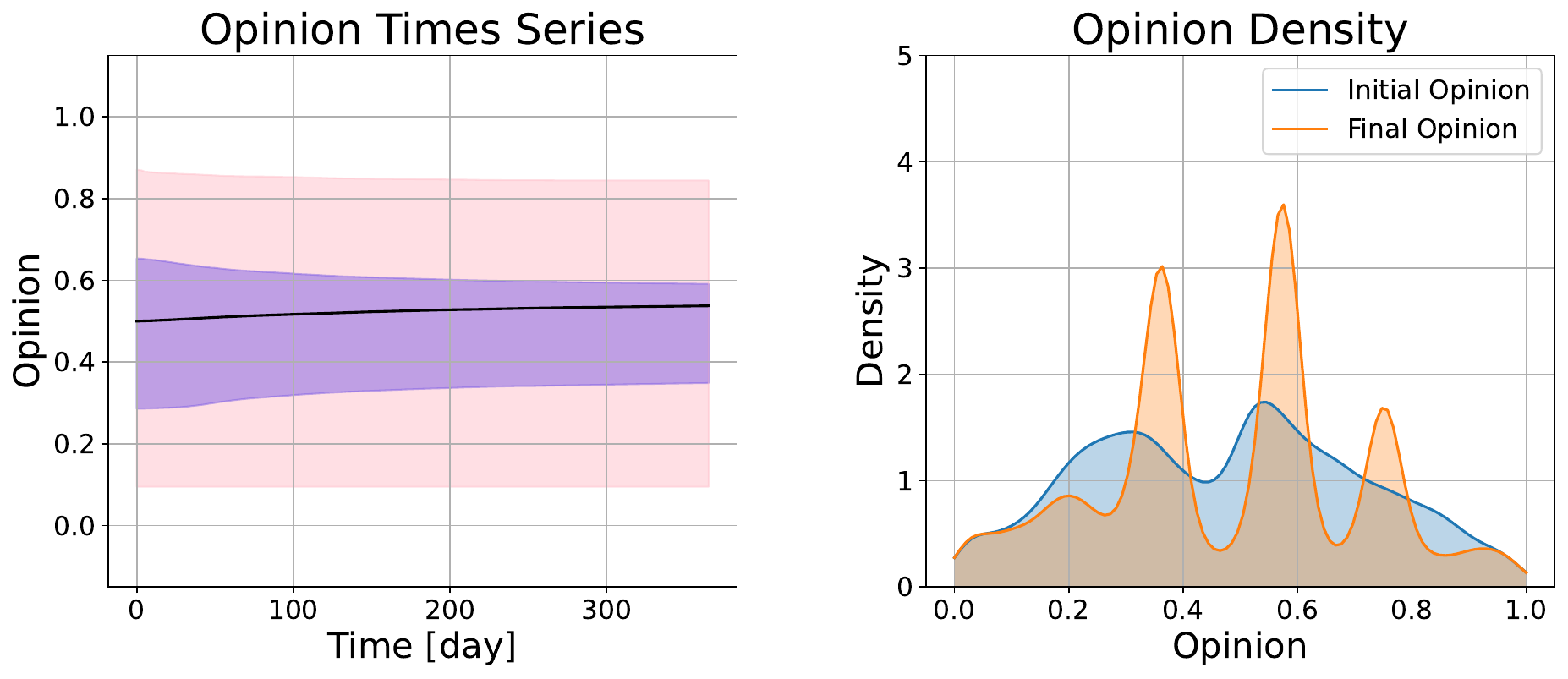}
    \caption{\textbf{Left:} Time series of opinions under natural dynamics (no agent intervention) on the U.S. election dataset. The purple band highlights the interquartile range (25th--75th percentiles), while the pink band indicates the 5th--95th percentile spread.\\[1ex]
    \textbf{Right:} Comparison of opinion densities at the start (blue) and at the end (orange) of the simulation under natural dynamics (no agent intervention) on the U.S. election dataset.}
    \label{fig:No_Agent_US_Election_sample}
\end{figure}

\begin{figure}[h]
    \centering
    \includegraphics[width=\textwidth]{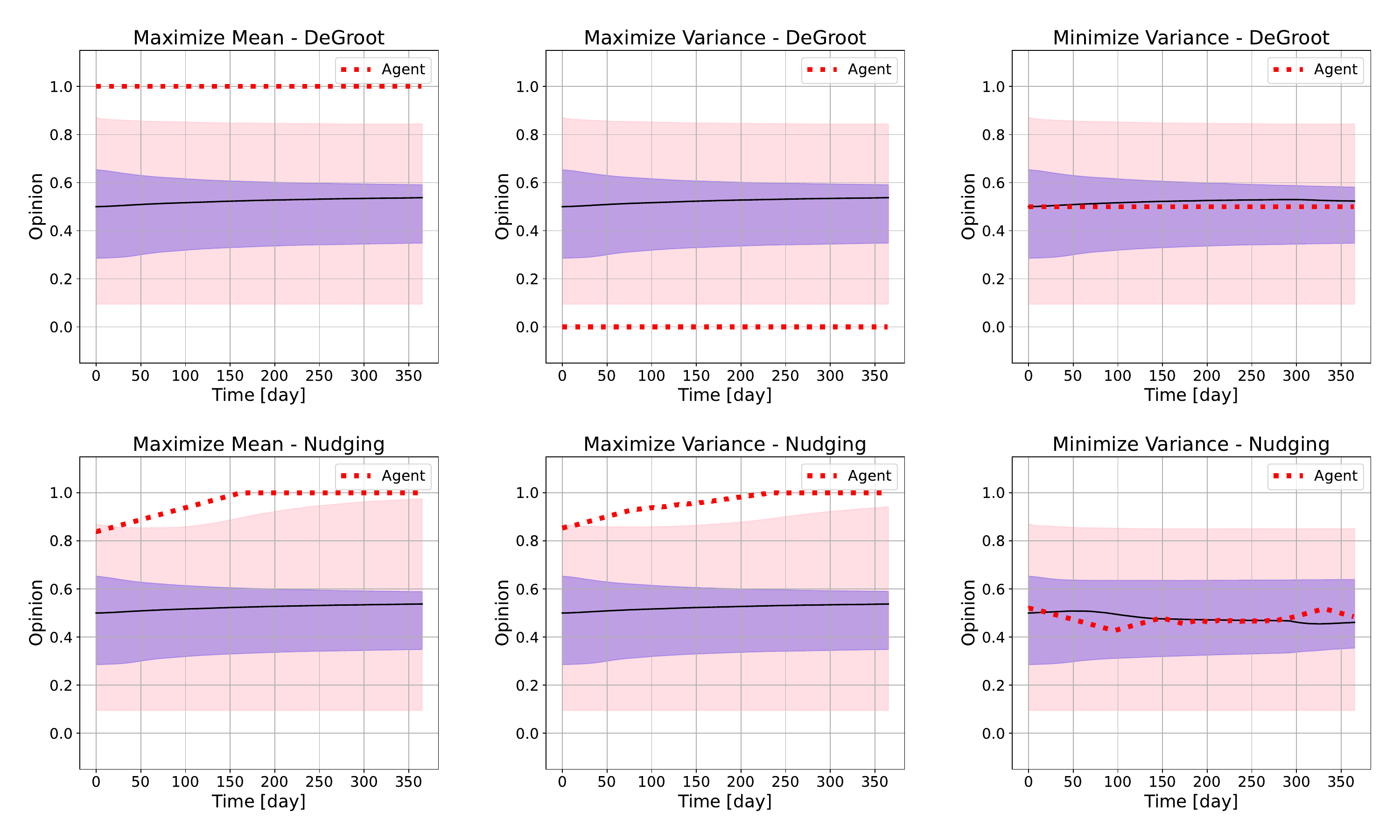}
    \caption{Time series depicting opinion distributions (shaded regions) and agent opinions (dashed lines) under DeGroot policy (top row) and nudging policy (bottom row) for different objectives on the U.S. election dataset. The objectives are maximizing mean (left), maximizing variance (middle), and minimizing variance (right). The purple region represents the 25th to 75th quantiles, while the pink region represents the 5th to 95th quantiles.}
    \label{fig:evolution_DeGroot_nudging_US_Election_sample}
\end{figure}

We further analyze how opinions evolve within the network using the U.S. election dataset. In Figure~\ref{fig:No_Agent_US_Election_sample}, we present time series and density plots of opinions under natural dynamics (i.e., without any agent intervention). Figure~\ref{fig:evolution_DeGroot_nudging_US_Election_sample} then illustrates how opinions evolve over time under the DeGroot and nudging policies.

Compared to the natural dynamics shown in Figure~\ref{fig:No_Agent_US_Election_sample}, the DeGroot policy produces minimal shifts in the opinion distributions because the agent’s opinion remains fixed. In contrast, the nudging policy employs a more strategic and adaptive adjustment of the agent’s opinion, leading to more effective persuasion across the network. For example, when the objective is to maximize the mean, the nudging agent steadily increases the 95th opinion quantile from 0.85 to 0.95. When maximizing variance, the nudging agent starts at 0.85 and gradually pushes the upper quantiles toward 1. However, the nudging effect is somewhat limited in this case, as a single agent is competing with the strong attractive dynamics among the network’s 30,000 nodes. For variance minimization, the nudging agent stabilizes near 0.5, effectively narrowing the interquartile range (the 25th to 75th percentiles).

\begin{figure}[h]
    \centering
    \includegraphics[width=\textwidth]{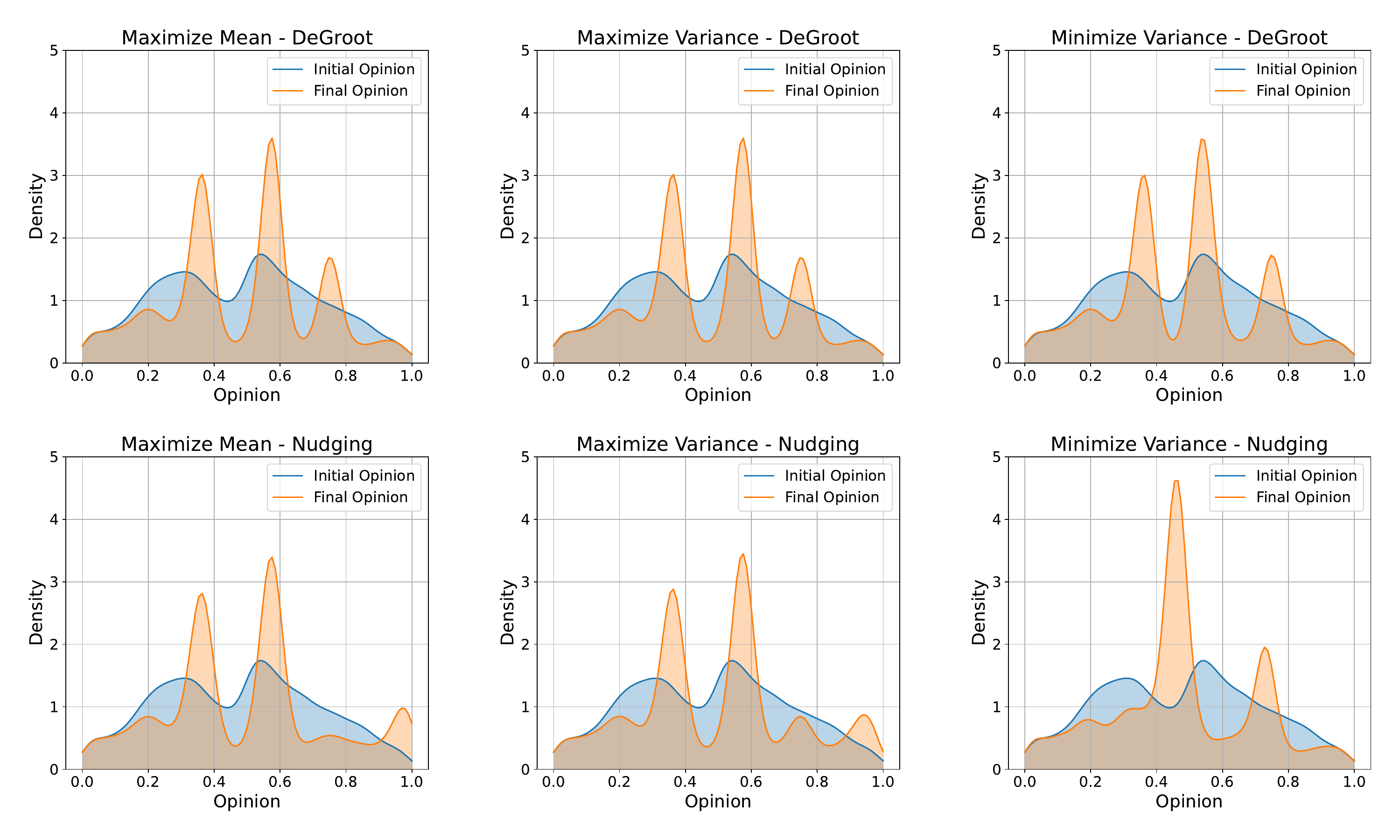}
    \caption{Initial (blue) and final (orange) opinion densities under DeGroot policy (top row) versus nudging policy (bottom row) for different objectives on the 2016 U.S. presidential election dataset. The objectives are to maximize mean (left), maximize variance (middle), and minimize variance (right).}
    \label{fig:density_DeGroot_nudging_US_Election_sample}
\end{figure}

We then compare the initial and final opinion densities under the two policies in Figure~\ref{fig:density_DeGroot_nudging_US_Election_sample}. The results show that the DeGroot agent causes only minor changes in the opinion distributions. Under the DeGroot policy, the final densities remain largely similar across all objectives—except for variance minimization, where the central mode shifts slightly toward 0.5. These final densities closely resemble those observed under natural dynamics in Figure~\ref{fig:No_Agent_US_Election_sample}, which exhibits three distinct modes (left, center, and right). As indicated by the bar plots in Figure~\ref{fig:objectives_DeGroot_nudging}, the DeGroot policy produces only a modest improvement when minimizing variance. In contrast, the nudging policy actively reshapes the opinion distributions: to maximize the mean, it diminishes the center-right density to form a distinct right mode at 0.95; to maximize variance, it polarizes the right opinion cluster into two modes at 0.75 and 0.95; and to minimize variance, it merges the left and center modes, resulting in a sharp spike at 0.45. These findings underscore the dynamic adaptability of the nudging policy compared to the static approach of the DeGroot policy.

A similar analysis for the Gilets Jaunes dataset is provided in \ref{sec:GJ}.

\subsubsection{Multi-Agent Nudging}
Our findings show that our nudging policy, even with a single agent, outperforms the DeGroot policy. We now extend this analysis to demonstrate that deploying multiple nudging agents enhances performance compared with a single agent. Specifically, we examine two multi-agent scenarios: one with ten agents, each targeting ten followers, and another with 100 agents, each targeting a single follower. Both scenarios maintain a fixed total targeting budget of 100 followers. It is worth noting that the DeGroot model does not benefit from a multi-agent setup, as all agents have the same static opinion over time, making it functionally equivalent to the single-agent case.

\begin{figure}[h]
    \centering
    \includegraphics[width=\textwidth]{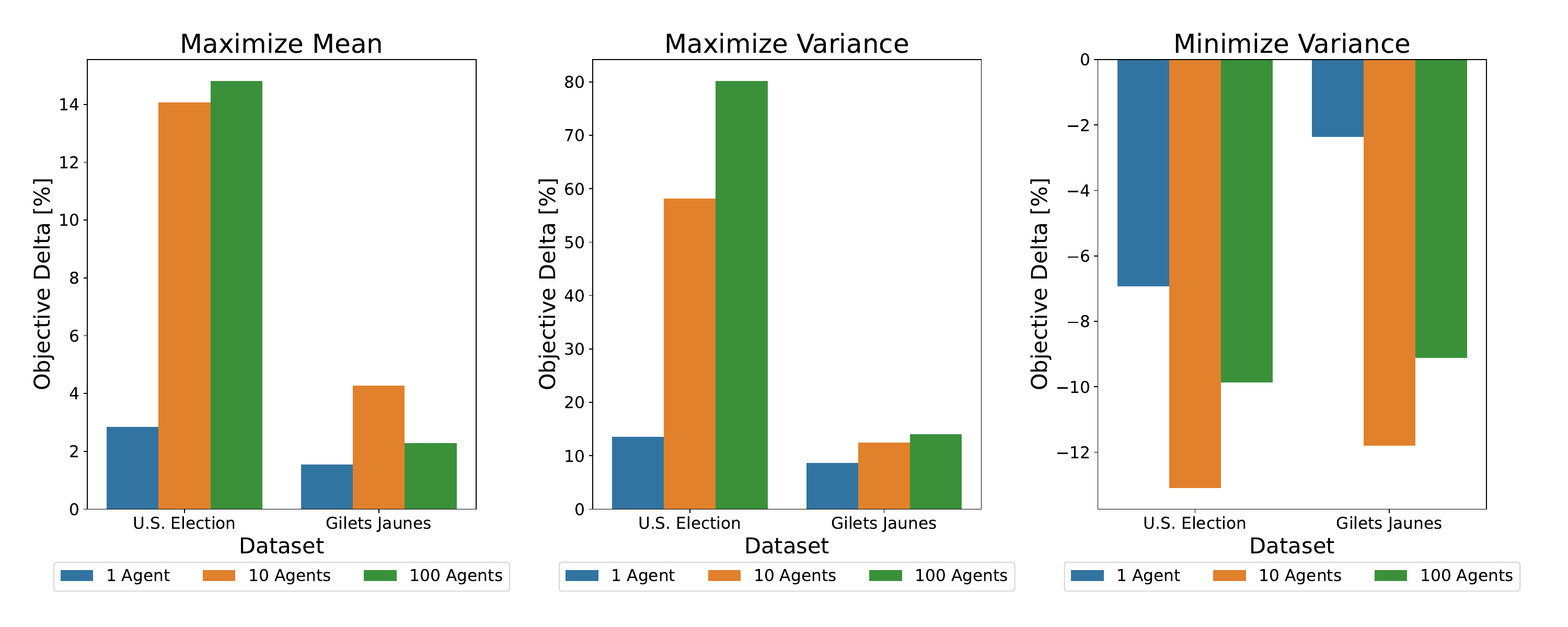}
    \caption{Bar plots comparing the change in final objective by dataset under varying numbers of nudging agents. The y-axis measures the percentage change of the final objective relative to the value with no agent. The objectives are maximizing mean (left), maximizing variance (middle), and minimizing variance (right).}
    \label{fig:objectives_nudging}
\end{figure}

Figure \ref{fig:objectives_nudging} shows that both the ten-agent and 100-agent scenarios surpass the single-agent scenario across all objectives. However, we observe that a larger number of agents does not necessarily yield better results. Specifically, in the case of minimizing variance, ten agents perform better than both one and 100 agents.  This is due to the greedy way in which targets are selected and contents are calculated, as stated in Proposition \ref{prop:adding_agents_suboptimal}. Our result suggests that in practice it is better to distribute the targets across a relatively small number of agents (relative to the number of targets).

\begin{figure}[h]
    \centering
    \includegraphics[width=\textwidth]{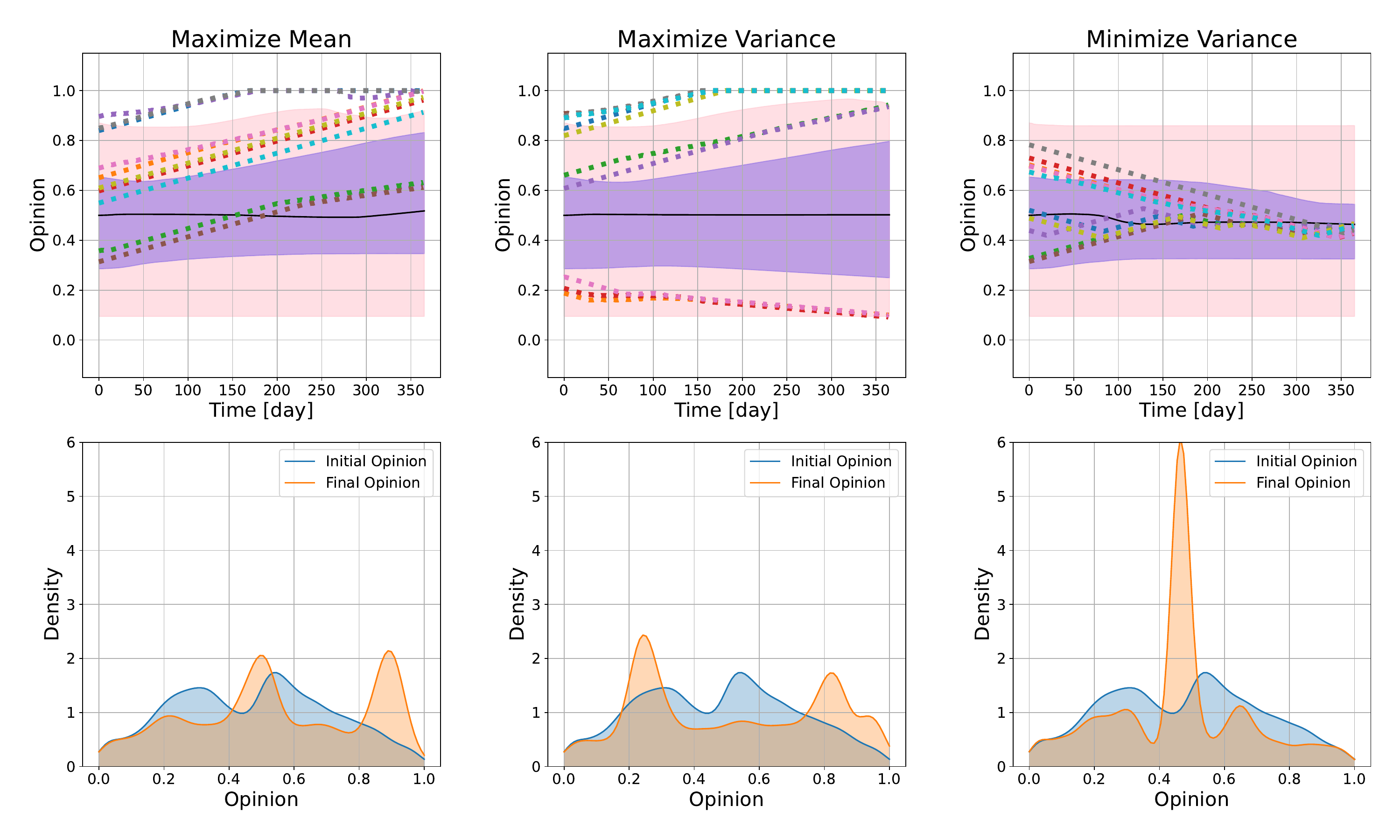}
    \caption{Opinion time series (top row) and density plots (bottom row) for different objectives under ten nudging agents on the U.S. election dataset. The time series depicts opinion distributions (shaded regions) and agent opinions (dashed lines). The purple region represents the 25th to 75th quantiles, while the pink region represents the 5th to 95th quantiles. Opinion densities are plotted for initial (blue) and final (orange) opinions. The objectives are to maximize mean (left), maximize variance (middle), and minimize variance (right).}
    \label{fig:nudging_10_agents_US_Election_sample}
\end{figure}

We visualize the opinion time series and densities in the ten-agent scenario depicted in Figure \ref{fig:nudging_10_agents_US_Election_sample} for the U.S. election dataset. For the multi-agent analysis of the Gilets Jaunes dataset, see \ref{sec:GJ}. The time series illustrates that the ten agents effectively and steadily nudge their respective follower clusters, influencing a broader span across the entire opinion spectrum. For maximizing the mean, the ten agents elevate the 75th to 95th quantiles, unlike the previous single agent who only affected the 95th quantile. Additionally, the ten agents reduce the left and center modes, which persist in the one-agent scenario, while also creating a taller right mode. In terms of maximizing the variance, we observe the significant formation of two modes at both extremes, a phenomenon not seen in the one-agent scenario. When minimizing the variance, the central mode at 0.5 becomes pronounced.

\begin{figure}[h]
    \centering
    \includegraphics[width=\textwidth]{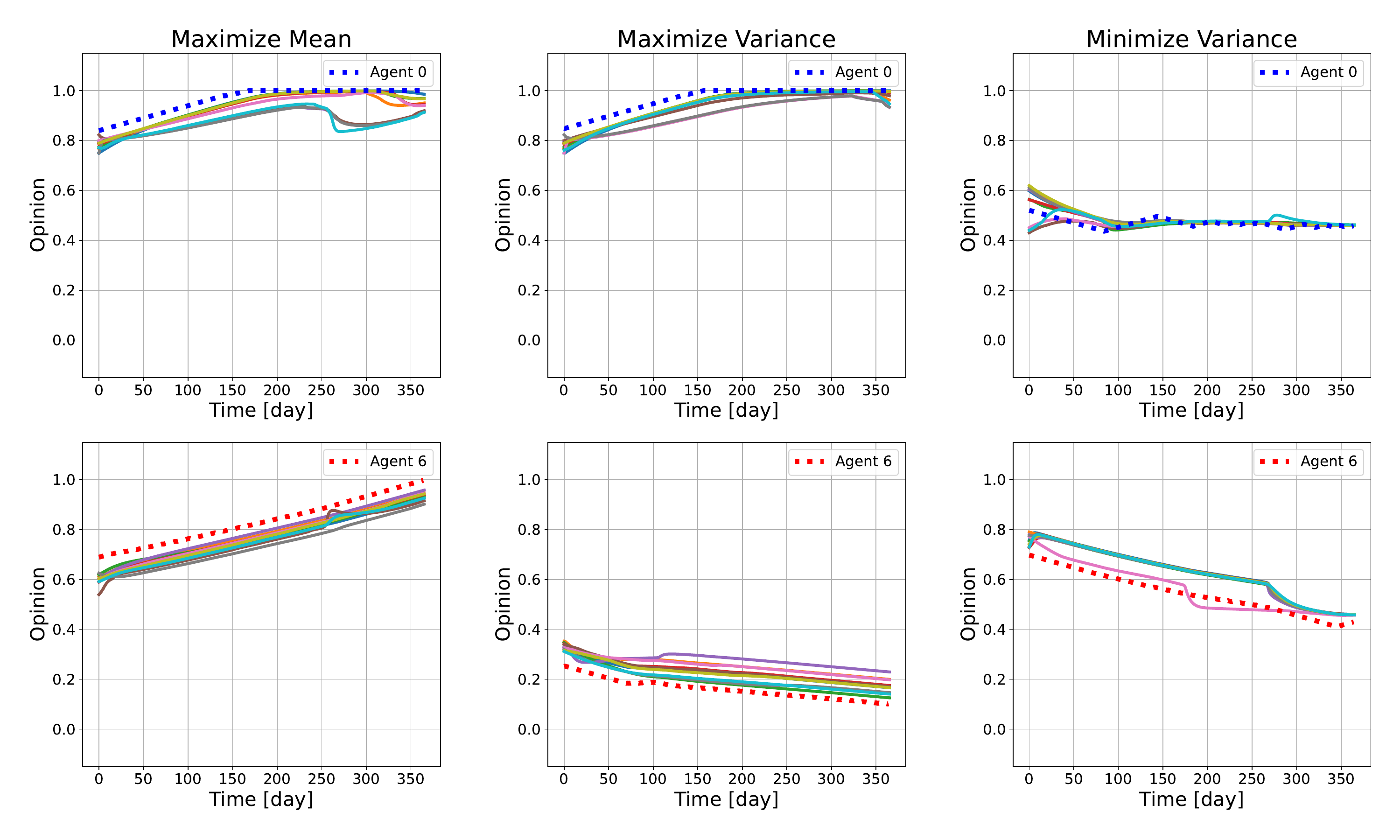}
    \caption{Opinion time series of the followers of agent 0 (top row) and agent 6 (bottom row) for different objectives under ten nudging agents on the U.S. election dataset. Each agent has ten followers with opinions shown in colored lines. Agent opinions are in dashed lines. The objectives are to maximize mean (left), maximize variance (middle), and minimize variance (right).}
    \label{fig:follower_evolution_nudging_10_agents_US_Election_sample}
\end{figure}

To get a better understanding of the agents' policies, we plot the opinions versus time for two agents, agent 0 and agent 6, along with the opinions of their followers (targets) for different objectives in the U.S. election dataset in Figure \ref{fig:follower_evolution_nudging_10_agents_US_Election_sample}.  From this we see the nudging policy in action.  Each agent's followers are initially within the confidence bound of the agent.  Then the content policy sees the agent pulling the followers either up or down, depending on the objective.  We see for the mean, both agents pull up their followers.  For maximizing the variance, the two agents target users slightly above and below the centrist opinion and then pull them towards the opposite extremes.  To minimize the variance, persuasion is achieved by drawing followers from extreme positions toward a more moderate, centrist stance.

\section{Content Creation with Large Language Models}\label{sec:llm}
So far, we have demonstrated how to determine targeting and content policies for agents in an influence campaign. The targeting policy, which dictates whom the agents should connect with in the network, is relatively straightforward to implement. In contrast, the content policy, which only assigns a numerical opinion to each piece of content, requires additional steps to be practically executed. Each opinion in the content policy must be converted into a piece of content the agent can post on social media in order to persuade individuals.  This can be accomplished through the use of large language models (LLMs) such at ChatGPT \citep{chatgpt1}, Claude \citep{claude}, Llama \citep{llama} and Gemini \citep{gemini}. These models allow for the controlled generation of text that has been shown to be persuasive on a variety of topics \citep{voelkel2023artificial, karinshak2023working, goldstein2024persuasive, costello_durably_2024, tessler2024ai}.  One can create a wide variety of text content by carefully prompting an LLM.  We now show the appropriate structure of the prompt for the content policy of an influence campaign.

LLMs allow one to create textual social media content with specific opinions on specific topics.  Among the variety of available LLMs, we utilize one of the latest versions of ChatGPT, known as GPT-4o-mini, which offers a balance of high-quality text generation and cost-efficiency \citep{chatgpt4o-mini}.  When querying ChatGPT, one can provide two types of text inputs: an instruction and a  prompt.  The instructions tell ChatGPT the general task it will be performing, while the prompt has data for a specific instance of the task.  For influence campaigns, we use the following instructions:
\newline
\texttt{You are going to help create content for a social
media account.  You will be asked to write persuasive content that has the given
opinion on the given topic.  Return only the text of the content.}

For the prompt, we first define some variables.  We let \pacman{topic} refer to the influence campaign's topic and \pacman{opinion} equal the numerical opinion of the content policy at a given time.  The lower and upper limits on the opinion range are \pacman{umin} and \pacman{umax}.   We let \pacman{content} be the type of social media content the agent will post on the social media platform.  For instance, if the influence campaign is run on Twitter, then we can set  \pacman{content} to be a tweet.  Given these variables, the prompt is:
\newline
\texttt{Write a \pacman{content} about "\pacman{topic}" that has an opinion of  \pacman{opinion}, where \pacman{umin} is total disagreement and \pacman{umax} is total agreement.
}
\newline
One can then insert values of the variables into the prompt to create the desired content.

We find that for the opinion range, it is better to use a scale centered at zero that ranges from negative to positive. ChatGPT is more consistently able to capture the correct valence, mapping negative opinions to content that does not support the topic and vice versa.  As an example, we set the topic to be ``I support Hillary Clinton for president.'' and the content to be a tweet.  We use two different opinion scales, one ranging from 0 to 100 and one ranging from -100 to 100.  We test three different opinion values, -80, 0, and 80 on the [-100,100] scale, which map to 10, 50, and 90 on the [0,100] scale through a simple linear transformation which aligns the extreme points and center of the scales.  The resulting tweets generated by ChatGPT are shown in Table \ref{table:opinion_scale}.  As can be seen, when we use the [-100,100] scale, the content with negative opinions do not support the topic, whereas for the [0,100] scale the content is all supportive.

\begin{table}[!hbt]
\centering
\begin{tabularx}{\textwidth}{|>{\centering\arraybackslash}p{2cm}|>{\centering\arraybackslash}p{2cm}|X|}
\hline
\textbf{Scale} & \textbf{Opinion} & \textbf{Text} \\
\hline
\rowcolor{red!25} $-100, 100$ & -80 & Honestly, I'm not really a fan of Hillary Clinton for president. \\
\hline
\rowcolor{red!25} $0, 100$ & 10 & Hillary Clinton has the experience, resilience, and vision to lead our nation forward. Her dedication to public service and commitment to equality make her the perfect choice for president! $\#$ImWithHer $\#$Hillary2016 \\
\hline
\rowcolor{blue!25} $-100, 100$ & 0 & Hillary Clinton is one of the candidates running for president. What are your thoughts on her policies and vision for the country? $\#$Election2016\\
\hline
\rowcolor{blue!25} $0, 100$ & 50 & Considering all the options for president, Hillary Clinton has valuable experience and policies worth contemplating. \#Election2016 $\#$ConsiderHillary \\
\hline
\rowcolor{cyan!25} $-100, 100$ & 80 & Hillary Clinton has the experience, dedication, and vision our country needs. She's a champion for equality and progress. Proud to support her for president! $\#$ImWithHer $\#$Clinton2016\\
\hline
\rowcolor{cyan!25} $0, 100$ & 90 &  Hillary Clinton's experience, dedication, and vision make her the ideal candidate for President. Her leadership can bring real change and progress! $\#$ImWithHer $\#$HillaryForPresident\\
\hline
\end{tabularx}
\caption{Tweets generated by ChatGPT with different opinion scales for the topic 
``I support Hillary Clinton for president.''}
\label{table:opinion_scale}
\end{table}

The choice of topic affects the focus of the content.  This is up to the discretion of those managing the influence campaign.  For instance, in the 2016 U.S. presidential election, one could make the topic focus on either of the two candidates and produces different types of content.  We show in Table \ref{table:candidates} tweets generated with the topic ``I support \textless{}candidate\textgreater{} for president'' where \textless{}candidate\textgreater{} is either Hillary Clinton or Donald Trump.  As can be seen, the tweet only mentions the candidate in the topic and makes no mention of their opponent.  This can be remedied by putting both candidates in the topic.  As an example, we show the tweets generated at different opinions for the topic ``I support Donald Trump for president and not Hillary Clinton in the 2016 election.'' in Table \ref{table:candidates_both}.  We see that now the tweets mention both candidates.  When running an influence campaign, one has discretion as to how to phrase the topic for the LLM.  The content policy allows for some freedom in the nature of the content.


\begin{table}[!hbt]
\centering
\begin{tabularx}{\textwidth}{|>{\centering\arraybackslash}p{3cm}|>{\centering\arraybackslash}p{2cm}|X|}
\hline
\textbf{\textless{}candidate\textgreater{}} & \textbf{Opinion} & \textbf{Text} \\
\hline
\rowcolor{red!25} Hillary Clinton & -80 & While Hillary Clinton has notable experience, I can't support her for president. We need fresh leadership and new ideas. $\#$NewDirection $\#$Election2016\\
\hline
\rowcolor{blue!25} Hillary Clinton & 0 & Considering all candidates, it's noteworthy to evaluate Hillary Clinton's experience and stance on key issues. $\#$Election2016\\
\hline
\rowcolor{cyan!25} Hillary Clinton & 80 & Hillary Clinton has the experience, resilience, and vision we need in a leader. She's fought for women's rights, healthcare, and fairness. She's ready to lead on day one. $\#$ImWithHer $\#$Hillary2016\\
\hline
\rowcolor{cyan!25} Donald Trump & -80 & While I respect differing opinions, I believe it's crucial to question and critically evaluate Donald Trump's suitability for the presidency based on his past actions and policies. $\#$NotMyPresident\\
\hline
\rowcolor{blue!25} Donald Trump & 0 & Balanced leadership is crucial for our nation’s future. Considering both positives and criticisms, it's essential to evaluate all candidates, including Donald Trump, with an open mind. Let's put America first by making informed choices. $\#$Election2016 $\#$CivicDuty\\
\hline
\rowcolor{red!25} Donald Trump & 80 & Donald Trump stands strong on traditional values and robust economic policies. His drive and leadership are what America needs to thrive. $\#$Trump2016 $\#$LeadershipMatters\\
\hline
\end{tabularx}
\caption{Tweets generated by ChatGPT for the topics ``I support \textless{}candidate\textgreater{} for president'' where \textless{}candidate\textgreater{} is either Hillary Clinton or Donald Trump.}
\label{table:candidates}
\end{table}

\begin{table}[!hbt]
\centering
\begin{tabularx}{\textwidth}{|>{\centering\arraybackslash}p{2cm}|X|}
\hline
\textbf{Opinion} & \textbf{Text} \\
\hline
\rowcolor{cyan!25} -80 & While I may disagree with both candidates, I am concerned that electing Trump could pose significant risks for our nation's future. $\#$Election2016 $\#$VoteWisely \\
\hline
\rowcolor{blue!25} 0 & Both Donald Trump and Hillary Clinton have their strengths and weaknesses. It's important to research and choose the candidate that aligns more closely with your values and priorities. $\#$Election2016 $\#$VoteSmart \\
\hline
\rowcolor{red!25} 80 & In the 2016 election, I support Donald Trump for president over Hillary Clinton. His bold vision for America's future \& commitment to shaking up the status quo resonate strongly with me! $\#$MAGA $\#$Election2016 \\
\hline
\end{tabularx}
\caption{Tweets generated by ChatGPT for the topic ``I support Donald Trump for president and not Hillary Clinton in the 2016 election.''}
\label{table:candidates_both}
\end{table}



\section{Conclusion}\label{sec:conclusion}
This study demonstrates the power of nudging-based influence policies in persuading users within online social networks while accounting for the bounded confidence effect, which limits the effectiveness of direct persuasion. By integrating control theory into the analysis of opinion dynamics, we developed a principled framework for designing influence campaigns that strategically guide audiences toward desired opinions. Our simulations on real Twitter networks confirm that these nudging policies not only shift average opinions but also modulate polarization—either increasing or reducing it based on campaign objectives.  

A key practical insight from our work is that these policies can be effectively implemented using AI, in particular, large language models.  This provides a scalable and adaptable approach for real-world influence operations. Moreover, the techniques we developed are language-agnostic, enabling their application across diverse cultural contexts, making them valuable tools for policymakers, strategists, and information operations practitioners.  

Our findings also highlight the risks of neglecting the bounded confidence effect when designing persuasion strategies. Under the classical DeGroot model, the most effective strategy involves static content policies that remain fixed and often present opinions starkly different from those of the target audience. However, our analysis reveals that when bounded confidence is considered, the optimal persuasion strategy is dynamic, adjusting over time to remain within the target audience’s persuadable range. This shift underscores the necessity of tailoring influence strategies to actual human opinion dynamics rather than relying on traditional models that assume unbounded receptivity to persuasion.  

By bridging mathematical modeling, optimal control, and generative AI, this study contributes both theoretical and practical advancements to the field of influence operations. As social media continues to evolve as a battleground for persuasion, understanding and responsibly leveraging realistic models of human behavior will be important for shaping public opinion in an increasingly complex digital landscape.  
  


\newpage
\pagenumbering{roman}
\bibliographystyle{aer}
\ifx\undefined\bysame
\newcommand{\bysame}{\leavevmode\hbox to\leftmargin{\hrulefill\,\,}}
\fi

\ECSwitch


\ECHead{Theorem Proofs and Twitter Network Results}
This e-companion includes the proofs of theorems, and the simulation results and analysis of the Gilets Jaunes Twitter dataset. The results of the U.S. election dataset are shown and discussed in the main article.

\section{Proofs}\label{sec:proof_theorems}

\subsection{Proof of Theorem \ref{thm:not_submodular_under_BC}}
\begin{figure}[H]
\centering
\includegraphics[width=\textwidth]{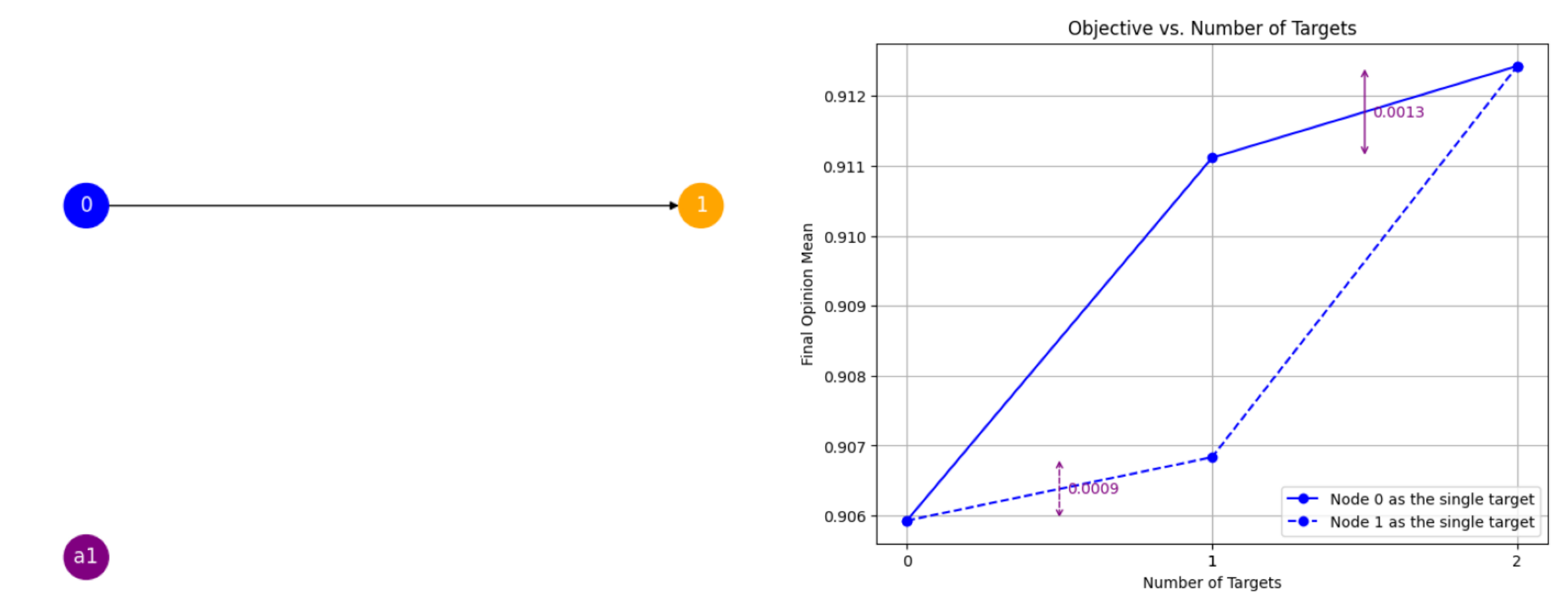}
\caption{Final opinion mean under bounded confidence opinion dynamics for a two-node network. The left panel depicts the network structure, where an agent with a fixed opinion of one interacts with two nodes. The right panel shows how the final opinion mean varies with the number of targets. In the single-target case, targeting node 0 (solid line) versus node 1 (dashed line) produces distinct outcomes; when targeting zero or both nodes, the results coincide.}
\label{fig:theorem_1_proof_mean_2_nodes}
\end{figure}

We provide examples to demonstrate that, under bounded confidence opinion dynamics, target selection is not submodular with respect to the number of targets. Consider the two-node network illustrated in Figure \ref{fig:theorem_1_proof_mean_2_nodes}. Our objective is to maximize the opinion mean at a final time $T$. One agent maintains a fixed opinion of one throughout the time period, tweeting once per unit time. The initial opinions of nodes 0 and 1 are 0.91 and 0.89, respectively, with tweet rates of 100 and one per unit time. The opinion shift function parameters are set to $\epsilon = 0.1$ and $\omega = 0.003$.

Let set \( A \) contain no target, and set \( B \) contain one target, node 0. Thus, \( A = \{\emptyset\} \), \( B = \{0\} \), and \( V = \{0, 1\} \), with \( A \subseteq B \subseteq V \). We now aim to show non-submodularity. Let \( r(\cdot) \) be the objective function, representing the final opinion mean as a function of the target set. Setting the final time \( T = 3 \), we find that \( r(A) = 0.9059 \) and \( r(B) = 0.9111 \). If we add one more target, node 1 to both target sets \( A \) and \( B \), we will have \( r(A \cup \{1\}) = 0.9068 \) and \( r(B \cup \{1\}) = 0.9124 \). Calculating the marginal gain, we observe:
\[
r(A \cup \{1\}) - r(A) < r(B \cup \{1\}) - r(B). \Halmos
\]
This result contradicts submodularity. Additionally, as shown on the right panel of Figure \ref{fig:theorem_1_proof_mean_2_nodes}, adding an additional target can result in a larger marginal objective value (the dashed line), indicating that adding more targets does not necessarily yield diminishing returns on the objective.

\begin{figure}[H]
\centering
\includegraphics[width=\textwidth]{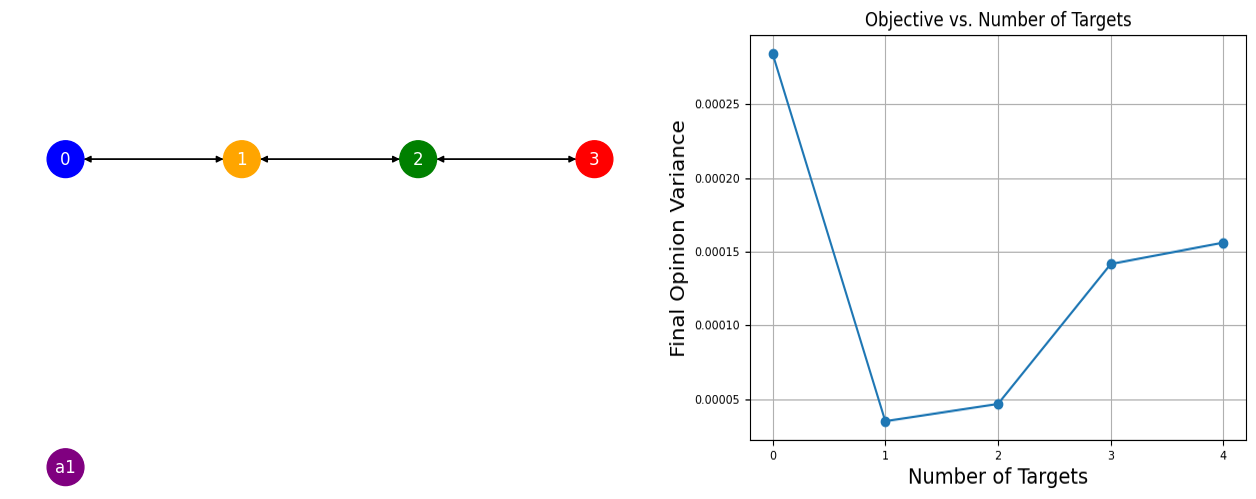}
\caption{Final opinion variance for a four-node path network (left) under different numbers of targets (right) within bounded confidence dynamics. A single agent maintains a fixed opinion of one over time. When targeting one node, the agent targets node 0. When targeting two nodes, the agent targets nodes 0 and 1. When targeting three nodes, the agent targets nodes 1, 2, and 3. When targeting four nodes, the agent targets the entire network.}
\label{fig:theorem_1_proof_variance}
\end{figure}

To demonstrate that the variance objective is also non-submodular, consider a four-node path network illustrated in the left panel of Figure \ref{fig:theorem_1_proof_variance}. The initial opinions of the nodes, in ascending order of node index, are 0.90, 0.92, 0.94, and 0.96, with respective tweet rates of one, one, 100, and one per unit time. Our objective is to maximize the final opinion variance by introducing one agent with a fixed opinion of one throughout the time period, tweeting ten times per unit time. The final time \( T = 10 \). The opinion shift function parameters are set to $\epsilon = 0.1$ and $\omega = 0.003$.

Let set \( A \) contain one target, node 0, and set \( B \) contain two targets, nodes 0 and 1. Thus, \( A = \{0\} \), \( B = \{0, 1\} \), and \( V = \{0,1,2,3\} \), with \( A \subseteq B \subseteq V \). We now aim to show non-submodularity. Let \( r(\cdot) \) be the objective function, representing the final opinion variance. We find that \( r(A) = 0.00003496 \) and \( r(B) = 0.00004669 \). If we add one more target, node 2 to both target sets \( A \) and \( B \), we will have \( r(A \cup \{2\}) = 0.00012825 \) and \( r(B \cup \{2\}) = 0.00014155 \). Calculating the marginal gain, we observe:
\[
r(A \cup \{2\}) - r(A) < r(B \cup \{2\}) - r(B). \Halmos
\]
Moreover, as shown in the right panel of Figure \ref{fig:theorem_1_proof_variance}, the variance is not even monotone with respect to the number of targets. 

For minimizing variance, the objective value is the negative of the objective when maximizing variance. Thus, minimizing variance is also non-submodular. In fact, one can refer to the theorem proved in \citep{hunter2022optimizing}, which states that under DeGroot dynamics, opinion variance is neither submodular nor monotone. Since DeGroot dynamics is a special case of the bounded confidence model with an infinite confidence interval, we conclude that under the bounded confidence model, the variance objective is not submodular with respect to the number of targets.

\subsection{Proof of Theorem \ref{thm:adding_agents_optimal}}

Consider the optimal control solution to the $k$-agent optimization problem (\ref{eq:optimal_control}), denoted by $(\boldsymbol{u}_1^*, \boldsymbol{u}_2^*, \ldots, \boldsymbol{u}_k^*, \boldsymbol{x}_1^*, \boldsymbol{x}_2^*, \ldots, \boldsymbol{x}_k^*)_k$. Here, each $\boldsymbol{u}_j^*$ represents the optimal opinion trajectory over time of the $j$-th agent, and $\boldsymbol{x}_j^*$ represents its corresponding target assignment. The subscript $k$ indicates that this tuple constitutes an admissible control for the $k$-agent optimization scenario, resulting in an optimal objective value:
\[
r^*_k = \max_{\boldsymbol{u}, \boldsymbol{x}} r(\boldsymbol{\theta}(T)).
\]

Our goal is to prove that adding one more agent (i.e., considering a $(k+1)$-agent scenario) cannot decrease the maximum achievable objective value; formally, we want to show:
\[
r^*_{k+1} \geq r^*_k.
\]

\textbf{Step 1 (Baseline Feasible Solution).} Observe first that the $k$-agent optimal solution directly provides a feasible solution for the $(k+1)$-agent problem by simply adding an inactive agent. Formally, define:
\[
(\boldsymbol{u}_1^*, \boldsymbol{u}_2^*, \ldots, \boldsymbol{u}_k^*, \boldsymbol{0}, \boldsymbol{x}_1^*, \boldsymbol{x}_2^*, \ldots, \boldsymbol{x}_k^*, \boldsymbol{0})_{k+1}
\]
as the baseline admissible control for the $(k+1)$-agent problem. Here, the $(k+1)$-th agent contributes zero opinion influence and targets no nodes, thereby achieving exactly the same objective value $r^*_k$.

\textbf{Step 2 (Improved Solution).} Starting from this baseline, we introduce a strategic improvement by activating the new $(k+1)$-th agent. Specifically, we consider shifting exactly one target node (without loss of generality, node 1) from its original agent assignment (say agent $k$) to the newly introduced $(k+1)$-th agent. We retain all other agents' targeting and opinion trajectories unchanged.

If the new $(k+1)$-th agent simply replicated the original agent's (agent $k$'s) opinion trajectory for node 1, the objective would remain $r^*_k$. However, we can improve this by cleverly positioning the $(k+1)$-th agent's opinion near node 1 at the final time step. Explicitly, assuming we want to maximize the opinion mean, we construct the following improved feasible solution for the $(k+1)$-agent scenario:
\begin{align*}
    (\boldsymbol{\hat{u}}_1, \boldsymbol{\hat{u}}_2, \ldots, \boldsymbol{\hat{u}}_k) &= (\boldsymbol{u}_1^*, \boldsymbol{u}_2^*, \ldots, \boldsymbol{u}_k^*), \\
    (\boldsymbol{\hat{x}}_1, \boldsymbol{\hat{x}}_2, \ldots, \boldsymbol{\hat{x}}_k) &= (\boldsymbol{x}_1^*, \boldsymbol{x}_2^*, \ldots, \boldsymbol{x}_k^* \setminus \{1\}), \\
    \hat{u}_{k+1}(t) &= 
    \begin{cases}
        u_k^*(t), & \text{for } t \in [0, T-\delta), \\
        \theta_1(t) + \epsilon, & \text{for } t \in [T-\delta, T],
    \end{cases} \\
    \hat{x}_{(k+1)i} &= 
    \begin{cases}
        1, & \text{if } i = 1, \\
        0, & \text{otherwise},
    \end{cases}
\end{align*}
where $\delta = 1/\sum_{(j,i)\in E}\lambda_{ji}$ is as defined in \citep{chen_shaping_2024}, and $\epsilon$ denotes the confidence interval in the bounded confidence model.

Thus, the newly introduced agent targets exactly node $1$ and at the final time interval strategically positions its opinion at the edge of node $1$'s confidence interval ($\theta_1 + \epsilon$), maximizing its impact.

\textbf{Step 3 (Evaluating the Improved Solution).} Denote the objective value corresponding to this new feasible solution as $\hat{r}_{k+1}$. Since only node $1$'s assignment is changed (from agent $k$ to agent $k+1$), we have:
\begin{align*}
    \hat{r}_{k+1} 
    &= r^*_k + \frac{\delta}{|V|}\cdot \lambda_{max}[f(\hat{u}_{k+1}(T-\delta)-\theta_1) - f(u_k^*(T-\delta)-\theta_1)] \\
    &= r^*_k + \frac{\delta}{|V|}\cdot \lambda_{max}[f(\epsilon) - f(u_k^*(T-\delta)-\theta_1)] \\
    &\geq r^*_k,
\end{align*}
since by definition of bounded confidence opinion dynamics, $f(\epsilon) \geq f(u_k^*(T-\delta)-\theta_1)$. Equality occurs only if $u_k^*(T-\delta)$ is exactly $\epsilon$ above $\theta_1$. Therefore, the constructed solution is guaranteed to perform at least as well as the $k$-agent optimal solution. Since the true optimal solution for $(k+1)$ agents, $r_{k+1}^*$, is by definition no worse than any feasible solution:
\[
r^*_{k+1} \geq \hat{r}_{k+1} \geq r^*_k. \quad\Halmos
\]

\textbf{Extension to Variance Objectives:} The same arguments hold if the objective is the opinion variance rather than the mean. The construction simply involves positioning $\hat{u}_{k+1}(T-\delta)$ strategically above or below $\theta_1$, depending on whether node 1 needs to be pulled toward or away from the opinion mean. Furthermore, strict improvement ($r^*_{k+1}>r^*_k$) can always be guaranteed if there exists at least one agent-target pair whose opinions are not exactly at the confidence interval distance $\epsilon$. Thus, the addition of another agent cannot decrease the achievable optimal objective and typically increases it.

\subsection{Proof of Proposition \ref{prop:adding_agents_suboptimal}}
We demonstrate this proposition through numerical examples. As illustrated in Figure \ref{fig:objectives_nudging} in the main article, employing ten agents consistently yields better objective values compared to a single agent. However, deploying 100 agents does not always outperform using ten agents. Let us assume \( k = 10 < k' = 100 \). We argue that in certain scenarios, the objective \( r_k \) for ten agents can exceed the objective \( r_{k'} \) for 100 agents.

In the U.S. election dataset, attempting to minimize variance with 100 agents results in a lower objective (i.e., more negative variance) than with ten agents. Specifically, we observe:
\[
r_k = -0.0433 > r_{k'} = -0.0449. \quad \Halmos
\]

Similarly, in the Gilets Jaunes dataset, when maximizing the mean, 100 agents produce worse outcomes than ten agents, where:
\[
r_k = 0.4987 > r_{k'} = 0.4892.
\Halmos
\]
Additionally, when minimizing variance in this dataset, we find:
\[
r_k = -0.0398 > r_{k'} = -0.0410.
\Halmos
\]

These results highlight a critical limitation of our greedy policy: increasing the number of agents does not necessarily lead to better outcomes. Therefore, numerical simulations play a pivotal role in determining the optimal number of agents in practice.

\section{Results of Twitter Networks}\label{sec:GJ}
Here we present the simulation results and discuss their implications within the Gilets Jaunes Twitter network. The results of our nudging-based policy is illustrated in Figures \ref{fig:evolution_DeGroot_nudging_Gilets_Jaunes_sample} through \ref{fig:follower_evolution_nudging_10_agents_Gilets_Jaunes_sample}, providing a detailed view of how the nudging policy influences opinion dynamics in this specific network context.

\subsection{DeGroot Policy versus Nudging Policy}
\begin{figure}[h]
    \centering
    \includegraphics[scale=0.4]{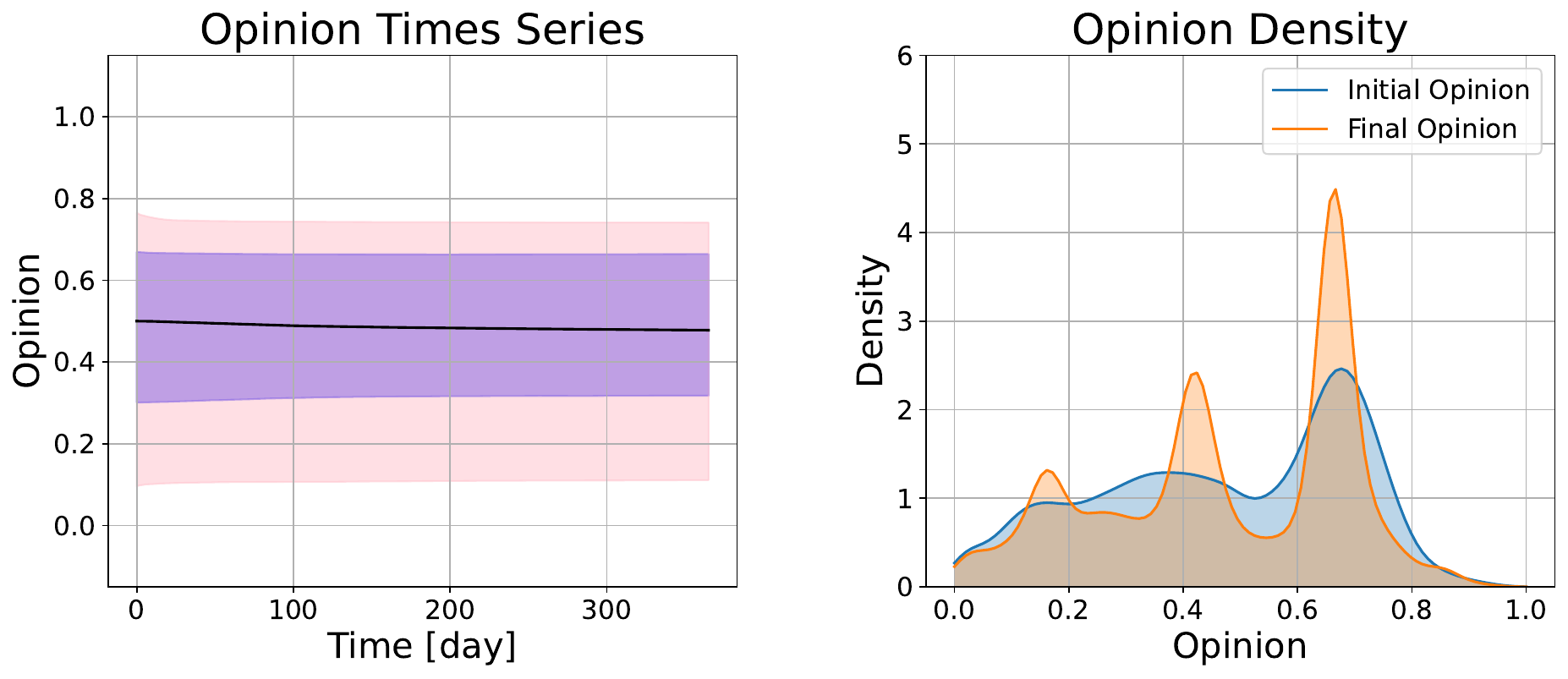}
    \caption{\textbf{Left:} Time series of opinions under natural dynamics (no agent intervention) on the Gilets Jaunes dataset. The purple band highlights the interquartile range (25th--75th percentiles), while the pink band indicates the 5th--95th percentile spread.\\[1ex]
    \textbf{Right:} Comparison of opinion densities at the start (blue) and at the end (orange) of the simulation under natural dynamics (no agent intervention) on the Gilets Jaunes dataset.}
    \label{fig:No_Agent_Gilets_Jaunes_sample}
\end{figure}

\begin{figure}[h]
    \centering
    \includegraphics[width=\textwidth]{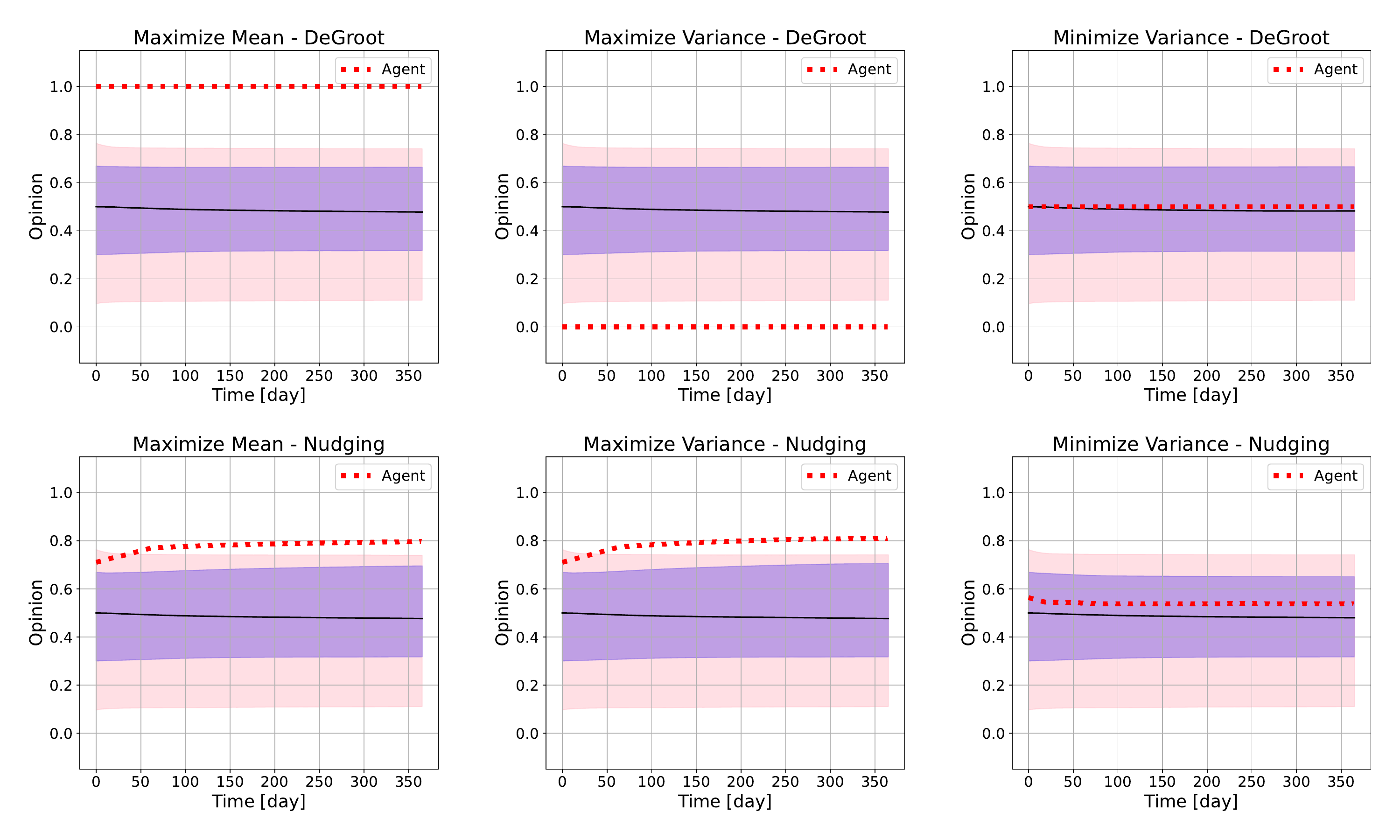}
    \caption{Time series depicting opinion distributions (shaded regions) and agent opinions (dashed lines) under DeGroot policy (top row) versus nudging policy (bottom row) for different objectives on the Gilets Jaunes dataset. The objectives are maximizing mean (left), maximizing variance (middle), and minimizing variance (right). The purple region represents the 25th to 75th quantiles, while the pink region represents the 5th to 95th quantiles.}
    \label{fig:evolution_DeGroot_nudging_Gilets_Jaunes_sample}
\end{figure}

\begin{figure}[h]
    \centering
    \includegraphics[width=\textwidth]{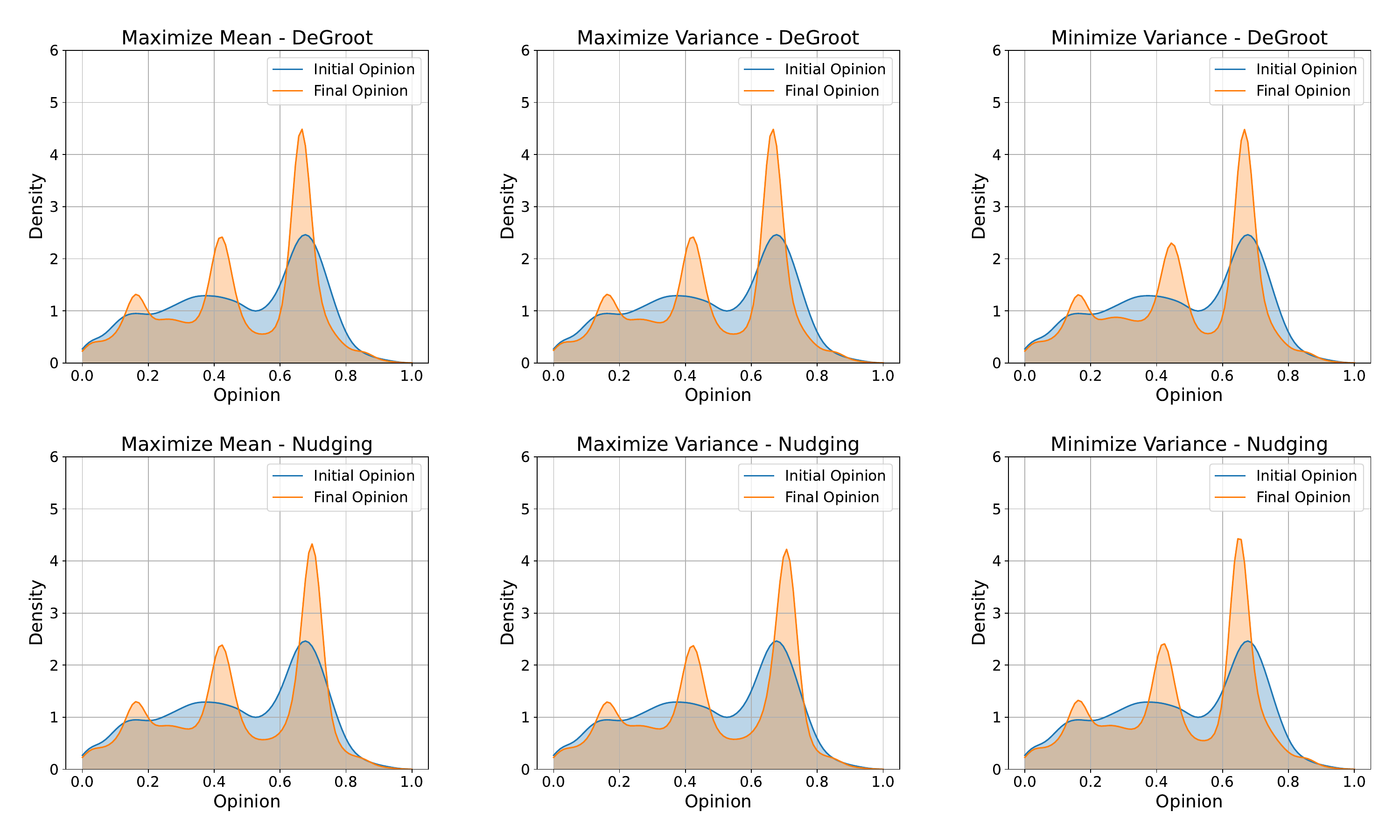}
    \caption{Initial (blue) and final (orange) opinion densities under DeGroot policy (top row) versus nudging policy (bottom row) for different objectives on the Gilets Jaunes dataset. The objectives are to maximize mean (left), maximize variance (middle), and minimize variance (right).}
    \label{fig:density_DeGroot_nudging_Gilets_Jaunes_sample}
\end{figure}

We begin by analyzing the effects of a single agent. As shown in the opinion time series in Figure \ref{fig:evolution_DeGroot_nudging_Gilets_Jaunes_sample} and the opinion density plot in Figure \ref{fig:density_DeGroot_nudging_Gilets_Jaunes_sample}, the DeGroot agent has minimal influence on shifting opinions. Across all objectives, the final opinion distributions closely match those observed under natural dynamics (i.e., without any agent intervention), as illustrated in Figure~\ref{fig:No_Agent_Gilets_Jaunes_sample}. In contrast, the nudging agent demonstrates a noticeable impact, significantly shifting opinions within the network. When aiming to maximize the mean opinion, the nudging agent effectively pushes the 75th quantile upward. Similarly, the agent increases the opinion variance by shifting the 75th quantile. To minimize variance, the density plot shows the right mode being pulled toward the center.

Although these shifts are not as pronounced as those observed in the U.S. election network, it is important to account for the fact that in the Gilets Jaunes network, the agent targets only a limited number of followers (see Table~\ref{table:number_targets}). Additionally, the Gilets Jaunes network shows a higher level of tweeting activity compared to the U.S. election network. This increased activity reinforces the natural opinion dynamics through stronger attraction forces, making it more challenging to exert external influence.

\subsection{Multi-Agent Nudging}
\begin{figure}[h]
    \centering
    \includegraphics[width=\textwidth]{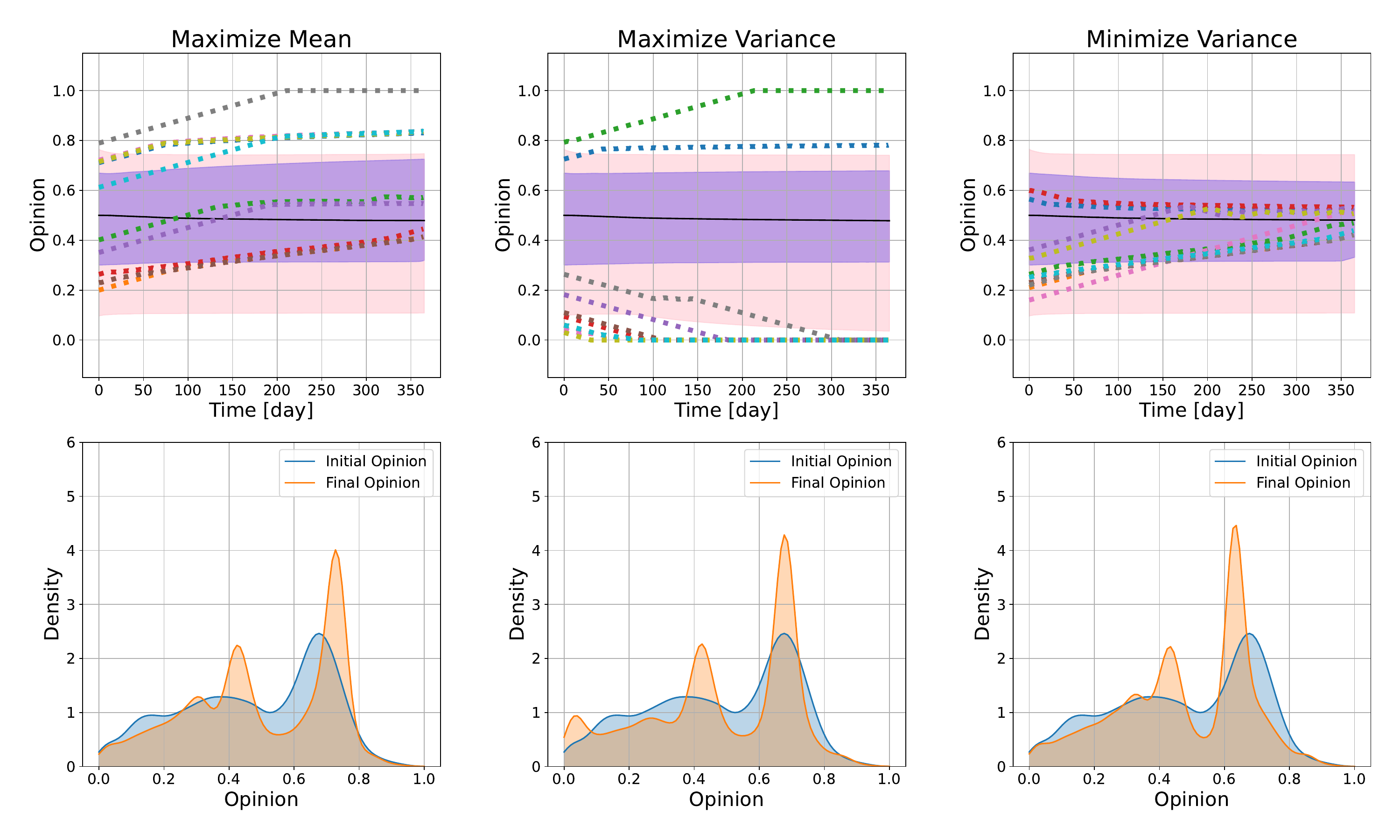}
    \caption{Opinion time series (top row) and density plots (bottom row) for different objectives under ten nudging agents on the Gilets Jaunes dataset. The time series depicts opinion distributions (shaded regions) and agent opinions (dashed lines). The purple region represents the 25th to 75th quantiles, while the pink region represents the 5th to 95th quantiles. Opinion densities are plotted for initial (blue) and final (orange) opinions. The objectives are to maximize mean (left), maximize variance (middle), and minimize variance (right).}
    \label{fig:nudging_10_agents_Gilets_Jaunes_sample}
\end{figure}

Multiple nudging agents outperform a single agent in terms of overall network influence. As shown in Figure \ref{fig:nudging_10_agents_Gilets_Jaunes_sample} which displays the opinion time series and density plots, employing ten nudging agents, each with ten followers, yields notable shifts in opinion dynamics. When the goal is to maximize the mean opinion, each agent targets specific follower clusters, collectively pulling the followers upward. This results in the amplification of the right mode at 0.7, while the left mode, initially at 0.15, is eliminated. For maximizing variance, the agents strategically divide the network, influencing both upper and lower follower clusters. This creates a distinct mode at 0.05. To minimize variance, the agents work to align their followers' opinions toward the center, around 0.5. The density plot clearly illustrates this effect, as the two modes are drawn closer together, converging toward 0.5.

\begin{figure}[h]
    \centering
    \includegraphics[width=\textwidth]{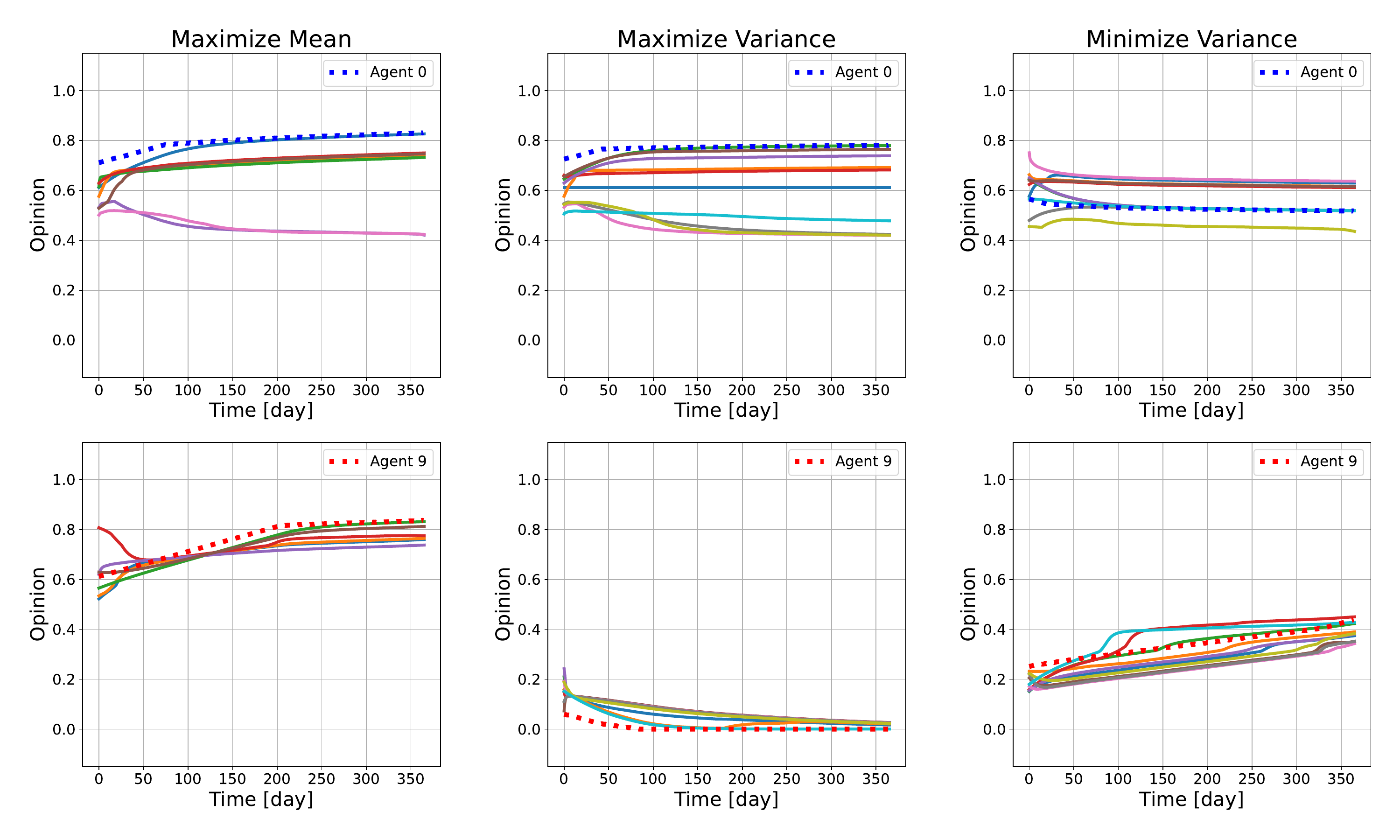}
    \caption{Opinion time series of the followers of agent 0 (top row) and agent 9 (bottom row) for different objectives under ten nudging agents on the Gilets Jaunes dataset. Each agent has a targeting budget of ten followers with opinions shown in colored lines. Agent opinions are in dashed lines. The objectives are to maximize mean (left), maximize variance (middle), and minimize variance (right).}
    \label{fig:follower_evolution_nudging_10_agents_Gilets_Jaunes_sample}
\end{figure}

We further examine follower behavior in Figure~\ref{fig:follower_evolution_nudging_10_agents_Gilets_Jaunes_sample}, focusing on agents 0 and 9 as representative examples. Each agent influences a distinct cluster within the opinion spectrum, gradually shifting their follower groups upward, downward, or toward the center. Notably, agents sometimes lose followers. For instance, when maximizing the mean, agent 0 loses two followers (highlighted in purple and pink). Initially, these followers are drawn toward the agent, but after the simulation period used in the targeting phase—where targets are selected using our greedy approach (Algorithm~\ref{algo:target}) with $T=30$—the purple and pink nodes exit the influence zone, and their opinions decay to 0.4. This limitation could be mitigated  by extending the simulation period to $T=365$ in the targeting phase, provided that sufficient computational resources are available.

%
%
%








\end{document}